\documentclass[twocolumn]{aastex61}
\usepackage{amsmath}
\usepackage[caption=false]{subfig}
\usepackage{graphicx}
\usepackage{epstopdf}
\usepackage{float}
\usepackage{subfig}
\usepackage{gensymb}
\usepackage{units}

\begin{document}

\title{Off-Axis Synchrotron Light Curves from Full-Time-Domain Moving-Mesh Simulations of Jets from Massive Stars}

\correspondingauthor{Xiaoyi Xie}
\email{xiaoyi@nyu.edu,macfadyen@nyu.edu}

\author[0000-0002-2798-6880]{Xiaoyi Xie}
\affil{Center for Cosmology and Particle Physics, Physics Department, New York University, 726 Broadway,
New York, NY 10003, USA}
\affil{Mathematical Sciences and STAG Research Centre, University of Southampton, Southampton SO17 1BJ, United Kingdom}

\author[0000-0002-0106-9013]{Andrew MacFadyen}
\affiliation{Center for Cosmology and Particle Physics, Physics Department, New York University, 726 Broadway,
New York, NY 10003, USA}



\begin{abstract}
  We present full-time-domain, moving-mesh, relativistic hydrodynamic simulations of jets launched from the center of a massive progenitor star and compute the resulting synchrotron light curves for observers at a range of viewing angles. We follow jet evolution from ignition inside the stellar center, propagation in the stellar envelope and breakout from the stellar surface, then through the coasting and deceleration phases. The jet compresses into a thin shell, sweeps up the circumstellar medium, and eventually enters the Newtonian phase. The jets naturally develop angular and radial structure due to hydro-dynamical interaction with surrounding gas. The calculated synchrotron light curves cover the observed temporal range of prompt to late afterglow phases of long gamma-ray bursts (LGRBs). The on-axis light curves exhibit an early emission pulse originating in shock-heated stellar material, followed by a shallow decay and a later steeper decay. The off-axis light curves rise earlier than previously expected for top-hat jet models -- on a time scale of
  seconds to minutes after jet breakout, and decay afterwards. Sometimes the off-axis light curves have later re-brightening components that can be contemporaneous with SNe Ic-bl emission. Our calculations may shed light on the structure of GRB outflows in the afterglow stage. The off-axis light curves from full-time-domain simulations advocate new light curve templates for the search of off-axis/orphan afterglows.
\end{abstract}




\section{Introduction} \label{sec:intro}
Gamma-ray bursts (GRBs) are extremely  energetic astrophysical
phenomena that emit bright multi-channel transient radiation. Long duration GRBs (LGRBs) are found to be associated with type
Ib/c supernova explosions (See e.g. \citealt{2006ARA&A..44..507W,2011AN....332..434M,2012grb..book..169H,2017AdAst2017E...5C} for recent reviews). 
The observations of GRBs have long revealed two
distinct phases, a prompt emission phase followed
by a long-duration afterglow phase. A longstanding physical model for GRBs is the fireball model, in which the prompt $\gamma$-ray emission comes from internal dissipation, and the broadband afterglow is produced by external shocks with the surrounding medium (See \citealt{1999PhR...314..575P,2004RvMP...76.1143P,2006RPPh...69.2259M,2015PhR...561....1K} for reviews). Many GRB models consider the dynamics and radiation from a ``top-hat jet'' : a uniform outflow with a well-defined sharp edge (e.g. \citealt{1997ApJ...487L...1R,1999ApJ...526..707P,1999ApJ...519L..17S,2000ApJ...541L...9K, 2000ApJ...529..151M, 2001grba.conf..312G}). Analytic studies and hydro dynamical simulations  utilizing top-hat jet models can reproduce the achromatic break observed in the afterglow light curves of GRBs -- the so called ``jet break''
(e.g. \citealt{1999ApJ...523L.121H,1999ApJ...522L..39S}). The typical
explanation for the jet break is that when the relativistic jet
decelerates, the observer starts to see the edge of the relativistic
jet \citep{1997ApJ...487L...1R,1999ApJ...525..737R,1999ApJ...519L..17S}.

A natural prediction of the off-axis light curves calculated from top-hat jet models is the existence of orphan afterglows (OAs). The prompt GRB emission is strongly suppressed for off-axis observers due to relativistic beaming. However afterglow emission can be observed by off-axis observers when the inverse lorentz factor of the emitting material is less than the angle to the observer.
Off-axis light curve templates inferred from top-hat jet models have been utilized to
calculate the detection rate of OAs in X-ray, Optical, and radio surveys (e.g. \citealt{2002ApJ...576..120T,2002ApJ...579..699N,2007A&A...461..115Z,2015A&A...578A..71G}).
As of yet, OAs with light curves predicted from top-hat jets have not been definitely detected. 

It is natural, however, to expect the angular profile of jet energy $\epsilon(\theta) = dE/d\Omega$ to decrease away from the jet axis as found in numerical simulations \citep{1999ApJ...524..262M,2001ApJ...550..410M, 2000ApJ...531L.119A,2003ApJ...586..356Z}. Various analytic jet angular structures have been proposed in the literature \citep{1998ApJ...499..301M,2002MNRAS.332..945R,2002ApJ...571..876Z,2003ApJ...591.1075K,2003ApJ...591.1086G}, including the universal structured jet model where $\epsilon(\theta)\propto \theta^{-2}$, and the Gaussian jet model $\epsilon(\theta)\propto \rm{exp}(-\theta^2/2\theta_c^2)$, where $\theta_c$ is a characteristic angular scale. 

Relativistic hydrodynamic simulations of GRB jets using a variety of initial conditions have been presented in the literature to model the prompt emission phase (e.g. \citealt{2009ApJ...700L..47L,2011ApJ...732...34L,2011ApJ...732...26M,2013ApJ...767...19L,2018MNRAS.478.4553D}) and the afterglow phase (e.g. \citealt{2001grba.conf..312G,2009ApJ...698.1261Z, 2010ApJ...722..235V,2012ApJ...751...57D,2018ApJ...863...32D,2018arXiv180305856G}) separately. \cite{2015ApJ...806..205D} used the moving-mesh code -- \texttt{JET} \citep{2013ApJ...775...87D} in two-dimensional spherical coordinates to follow the jet from deep inside the star all the way into the afterglow phase, and calculated synchrotron light curves for on-axis observers. In this work, we present similar full-time-domain (FTD) numerical simulations that cover the life-cycle of jets from deep inside the star all the way to the Newtonian phase, and calculate light curves for observers at a range of off-axis viewing angles. Collisions between the injected jet material and the stellar envelope results the formation of an ultra-relativistic outflow (i.e. the jet). We locate the photospheric position as the simulations evolve in time and calculate synchrotron radiation from the optically thin regions of the outflow. The resulting light curves cover very early phases of synchrotron emission from jet expansion. There is an initial pulsed emission for on-axis light curves which mainly comes from shocked stellar material. This may shed light on the emitting source of observed long, smooth, and single-pulsed GRBs (see \citealt{2016ApJ...822...63B,2018ApJ...859..163H} for examples). The off-axis light curve rises very early on which differs from the late rise-ups found for top-hat Blandford-Mckee (BM) jet models. 
In Section \ref{sec:dynamics}, we demonstrate the numerical improvements utilized in the simulations. We have incorporated an effective adaptive mesh refinement (AMR) scheme into the moving-mesh code - \texttt{JET} \citep{2013ApJ...775...87D}. We compare the AMR-enhanced moving-mesh code and the Eulerian AMR code -- \texttt{RAM} in Section \ref{sec:bm}. In this section, we also discuss the classical top-hat BM hydrodynamic simulations and the associated afterglow light curves. In Section \ref{sec:star}, we present the FTD dynamical evolution of jets breaking out of a stellar progenitor and discuss the emerged jet structure. In Section \ref{sec:star_rad}, we present on- and off-axis synchrotron light curves directly calculated from our simulations. Implications from these light curve features are made. We conclude our findings in Section \ref{sec:discussion}.

\section{Numerical Method}\label{sec:dynamics}

The simulations are performed with a 2D spherical axi-symmetric relativistic hydrodynamic (RHD)
 moving-mesh code-- {\texttt{JET}} \citep{2013ApJ...775...87D}.
The code numerically integrates the following equations:
\begin{eqnarray}
  \partial_\mu (\rho u^\mu) = S_D\, , \label{eq:hydro1}\\
  \partial_\mu ( \rho h u^\mu u^\nu + P g^{\mu\nu}) = S^{\nu}\,,\label{eq:hydro2}
\end{eqnarray}
where $\rho$ is proper density, $\rho h = \rho + \epsilon + P$ is
enthalpy density, $P$ is pressure, $\epsilon$ is internal energy
density, and $u^{\mu}$ is the four-velocity, where the speed of light $c$ is set to one. The equations are solved
in two dimensional spherical coordinates assuming axisymmetry. The
source terms $S_D$ and $S^\nu$ are used to model the injection of
mass, momentum, and energy on small scales by the central engine.
We employ RC equation of state (EOS) in our simulations which matches the exact Synge equation within an accuracy of $0.8\%$
\citep{2006ApJS..166..410R}. We express the specific enthalpy as a
 function of $\Theta=P/\rho c^2$. and utilize Newton-Raphson iteration to  find the root of $\Theta$ based on the values of conservative variables. The new primitive variables are then calculated accordingly. Detailed procedures are covered in 
 \citealt{2012ApJ...746..122D}. Different EOS could be adopted in simulations by changing the expression of $h(\Theta)$.  Equation \ref{eq:gas} - \ref{eq:ryu} corresponds to the specific enthalpy function subject to ID (ideal
 gas law)  EOS, TM EOS \citep{2005ApJS..160..199M}, and RC EOS \citep{2006ApJS..166..410R},
 respectively:

\begin{eqnarray}
  h(\Theta) &=& 1 + \frac{\Gamma}{\Gamma-1}\Theta \label{eq:gas}\,, \\
  h(\Theta) &=& \frac{5}{2}\Theta + \frac{3}{2}\sqrt{\Theta^2 + \frac{4}{9}} \label{eq:tm}\,,\\
  h(\Theta) &=& 2\frac{6\Theta^2 + 4\Theta + 1}{3\Theta + 2} .\label{eq:ryu}
\end{eqnarray}

We use the HLLC Riemann solver \citep{2005MNRAS.364..126M} and move the radial numerical cell faces at
the local contact discontinuity (CD) velocity. In the vicinity of the shock front, high resolution along the radial direction is required to fully resolve the dynamics of the relativistic structures in the outflow. The AMR scheme, which actively refine cells in the relativistic region and derefine cells within non-relativistic gas, are implemented within Eulerian RHD code frame (e.g. \citealt{2000ApJS..131..273F,2006ApJS..164..255Z,2012ApJ...746..122D}). The time step size is, however, limited by the finest cells with high velocity. The moving-mesh technique, with each cell moving at approximate the local CD velocity, can enlarge the time step (see e.g. \citealt{2011ApJS..197...15D,2013ApJ...775...87D}). The combination of these two techniques in principle allows accurate simulations of relativistic jets in efficiency. In this work, we incorporate a robust AMR scheme into the moving-mesh
code. We define an approximate
numerical second derivative of fluid variables as a measurement of error:
\begin{equation}\label{eq:error}
  E_i = \frac{|u_{i+2}-2u_{i}+u_{i-2}|}{|u_{i+2}-u_i| + |u_i-u_{i-2}| + \delta (|u_{i+2}|+2|u_i|+|u_{i-2}|)},
\end{equation}
which is utilized in empirical AMR schemes (see e.g. \citealt{2000ApJS..131..273F,2006ApJS..164..255Z}). By default, we set the adjustable parameter $\delta$ in the denominator to 0.01. The spherical domain is evenly distributed in the angular direction with $\rm{N_\theta}\,=\,
320$ radial tracks, yielding an angular resolution of $0.28^{\degree}$. The radial tracks shear with the radial velocity of the gas. Each radial track moves independently, behaving essentially as 1D Lagrangian grid \citep{2013ApJ...775...87D}. In each time step, the cell along a radial track with the maximum measurement of error $E_{i,\rm{max}}$ will be refined if $E_{i,\rm{max}} > 0.9$.  The cells with $E_i<0.002$ will be considered to be derefined. The final choice of the cell to be
derefined, for each radial track, is the one that has the smallest time step. The time step of each cell is estimated according to the CFL
condition $\Delta t< \rm{cfl} \frac{\Delta r}{\rm{max}(v_r-w)}$, where $\rm{cfl}$ is a constant, $\Delta r$ is the radial cell length, $v_r$ is the characteristic wave speed calculated at the cell face, $w$ is the radial velocity of the cell face (i.e. the local CD velocity). Initially, the spherical domain is logarithmically spaced in the radial
direction, and the aspect ratio of each cell $S=\frac{r\Delta\theta}{\Delta r}$ is of order one. For relativistic explosions, the dynamical scale: $\Delta r/r < 1/16\Gamma^2$ sets the desired radial resolution of the relativistic shell.
This gives us an estimation of the required aspect ratio in order to resolve the relativistic shell: $S > 16\Gamma^2\Delta \theta\sim
784\,\Gamma_{2}^2,\ \Gamma_2=\frac{\Gamma}{100}$. This criteria has been incorporated into the AMR scheme to determine whether or not to derefine the high-resolution cells in the ultra-relativistic region. With cell faces moving radially at the local CD velocity, the aspect ratio of each cell is adjusting dynamically. The simulations performed in this study have maximum Lorentz factor around $100$. The maximum dynamic aspect ratio reaches about 450.
Also, we find the combined scheme is able to accurately simulate an ultra-thin relativistic jet with a Lorentz factor up to $10^4$. Compared with the traditional Eulerian AMR scheme, the AMR-enhanced moving-mesh scheme delivers accuracy in efficiency.



\section{Top-hat Blandford-McKee Models}\label{sec:bm}
\subsection{Code comparison}
We perform standard top-hat Blandford-Mckee (BM) \citep{1976PhFl...19.1130B} simulations to check the robustness of the new
AMR-enhanced moving-mesh scheme. For the BM solution, the Lorentz factor of the shock front $\Gamma$,  the Lorentz factor of the fluid $\gamma$, the elapsed time $t$, and the jet radius $R$ follow the relations (the speed of light $c$ is set to
one):
\begin{eqnarray}
  \Gamma &=& \sqrt{2}\gamma, l=(E_{\rm{iso}}/\rho_0)^{1/3}\,,\\
  t &=& l(\Gamma/\sqrt{17/8\pi})^{-2/3} \label{eq:bm_t}\,,\\
  R &=& (1-1/(8 \Gamma^2))t\label{eq:bm_r}.
\end{eqnarray}

\begin{figure}[!ht]
  \centering
  \includegraphics[width=0.45\textwidth]{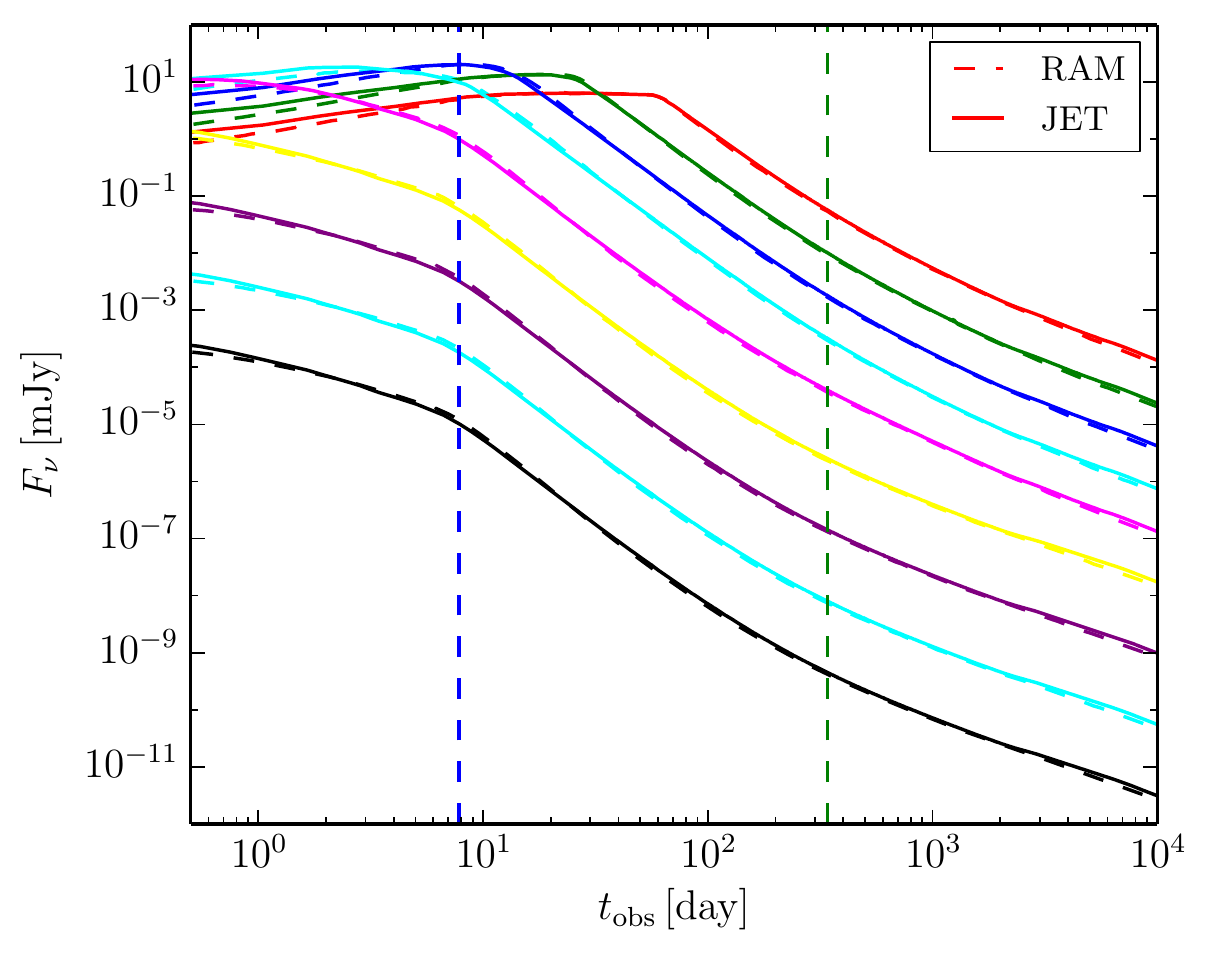}
  \caption{Code comparison for on-axis light curves calculated from top-hat Blandford-McKee simulations. 
    Solid lines represent light curves produced from the top-hat BM simulation using our AMR-enhanced moving-mesh code - \texttt{JET}. Dashed lines
    represent results from the same model performed with the Eulerian AMR code -- \texttt{RAM} (taken from  \citealt{2009ApJ...698.1261Z}). The flux density for
    various frequencies are plotted: $10^9\,\rm{Hz}$ (red),
    $10^{10}\,\rm{Hz}$ (green),$10^{11}\,\rm{Hz}$ (blue), $10^{12}\,\rm{Hz}$ (cyan),
    $10^{13}\,\rm{Hz}$ (magenta), $10^{14}\,\rm{Hz}$ (yellow), $10^{15}\,\rm{Hz}$ (purple),
    $10^{16}\,\rm{Hz}$ (aqua), and $10^{17}\,\rm{Hz}$ (black). The vertical dotted line (at  7.9 and 340 day) represents the jet break and the Newtonian transition time, respectively \citep{2009ApJ...698.1261Z}.}\label{fig:wq_compare}
\end{figure}

The half opening angle of the top-hat BM jet ($\theta_{\rm{jet}}$) is set to $\theta_{\rm{jet}}=0.2$. The isotropic equivalent energy is $E_{\rm{iso}} = 10^{53}\,\rm{erg}$. The uniform ambient density $\rho_0$ is set to 1 proton per cubic centimeter, and the pressure
is set to $P_0=10^{-10}\rho_0$ following previous Eulerian AMR simulations \citep{2009ApJ...698.1261Z,2010ApJ...722..235V}. The
hydrodynamic simulation starts at the moment when the initial
Lorentz factor of the fluid just behind the shock is $\gamma=20$. 
In Figure \ref{fig:wq_compare}, we demonstrate that the on-axis synchrotron light curves calculated from our BM simulation match well with results (as presented in \citealt{2009ApJ...698.1261Z}) from simulations performed with the Eulerian AMR code - \texttt{RAM} \citep{2006ApJS..164..255Z}. We calculate the broadband synchrotron radiation light curves utilizing a well-tested synchrotron radiation code (\citealt{2009ApJ...698.1261Z,2010ApJ...722..235V}, see also \citealt{1998ApJ...497L..17S,2012ApJ...746..122D}). The same radiation parameter values are used in both calculations: 
the electron equipartition factor $\epsilon_e = 0.1$, the magnetic equipartition factor $\epsilon_B = 0.1$, and the energy power-law index of
relativistic electron $p=2.5$. The flux distance scaling is set to $ 1/{4\pi d_{28}^2}, d_{28}=10^{28}\,\rm{cm}$ 
as in \cite{2009ApJ...698.1261Z}.
We also adopt the same EOS (TM-EOS) in this simulation for a fair comparison.

\begin{figure}[!ht]
  \centering
  \includegraphics[width=0.45\textwidth]{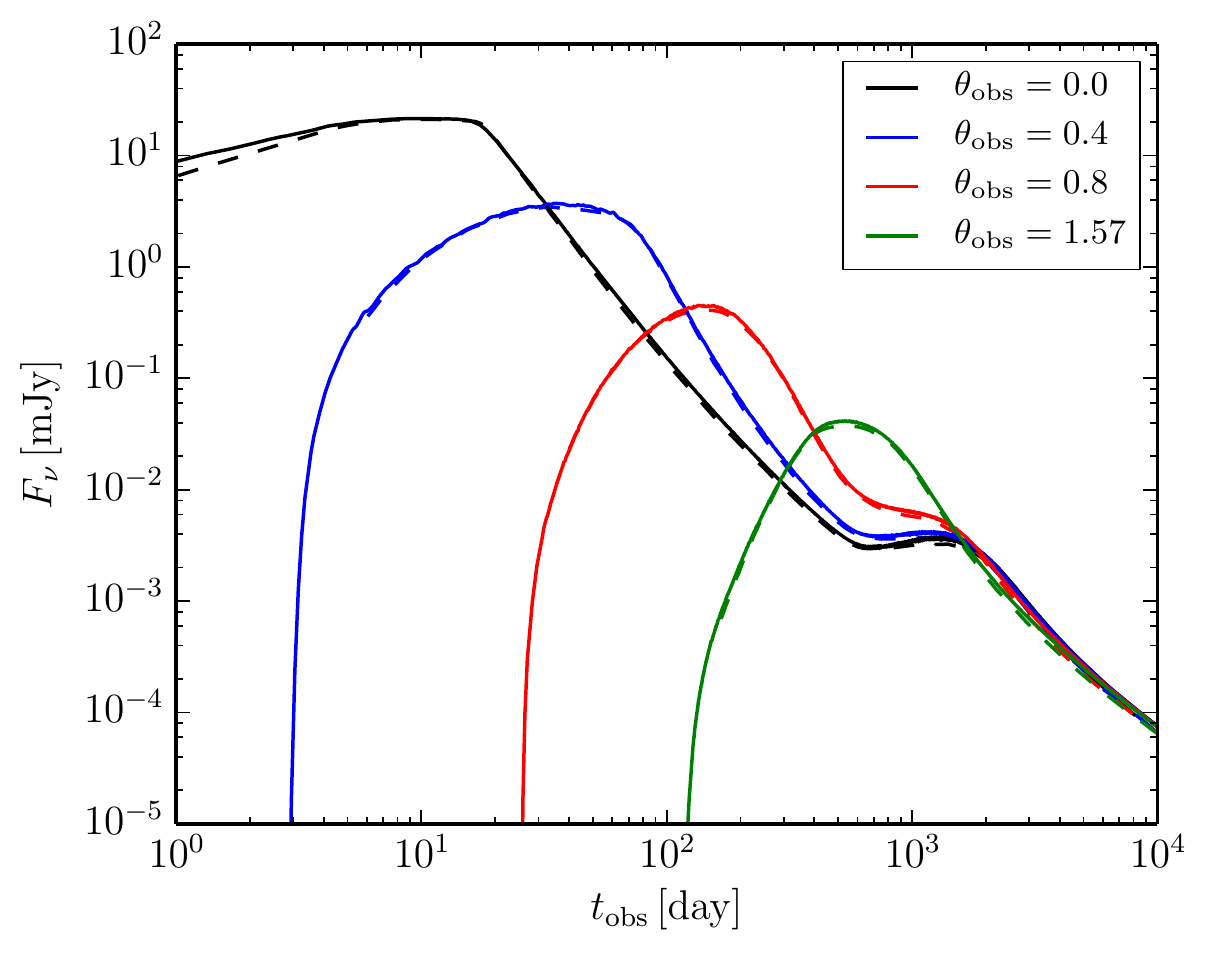}
  \caption{Code comparison for off-axis BM light curves between the
    AMR-enhanced moving-mesh code -- \texttt{JET} (solid line) and the Eulerian AMR code 
    -- \texttt{RAM} (dashed line). The light curves are calculated from top-hat BM simulations at Radio frequency $8.46\,\rm{GHz}$.
    Light curves with increasing viewing angle are presented from top to bottom. The off-axis light curves from \texttt{RAM} code are taken from \cite{2010ApJ...722..235V}.}\label{fig:hk_compare}
\end{figure}

For off-axis light curves, we compare with results from 
\cite{2010ApJ...722..235V}.  
As shown in Figure \ref{fig:hk_compare}, off-axis light curves
from our BM simulation (solid lines) match well with results from \citet{2010ApJ...722..235V}. One advantage of our new scheme is that we're able to perform accurate top-hat BM simulations at an earlier time with Lorentz factor of hundreds instead of tens.

Observational results indicate that GRB afterglow comes from ultra-relativistic jets with inferred
Lorentz factor $\sim 100$. Numerical simulations with initial Lorentz factor $\sim 100$ are thus necessary to fully capture early afterglow emission. Here, we perform two top-hat BM simulations with different initial Lorentz factors: (1) $t_0=
4.37\times 10^{6}\,\rm{s},\,\gamma_0=100,\,R_0=1.31\times
10^{17}\,\rm{cm}\ \rm{and}\ \theta_{\rm{jet}}=0.2$ (denoted as BM-J0.2-G100 simulation). (2)
$t_0=1.28\times 10^{7}\,\rm{s},\,\gamma_0=20,\,R_0=3.83\times
10^{17}\,\rm{cm}\ \rm{and}\ \theta_{\rm{jet}}=0.2$ (denoted as BM-J0.2-G20 simulation), respectively. The RC-EOS is adopted for both BM simulations (hereafter, all the simulations are performed
with RC-EOS). The synchrotron radiation parameter values are listed in
Table \ref{tab:sync}. The on- and off-axis synchrotron light curves at frequency $10^{17}\,\rm{Hz}$ from the BM-J0.2-G20 simulation  and from
the BM-J0.2-G100 simulation are over-plotted in Figure \ref{fig:lc_g20_g100_offaxis}.
The early on-axis light curve calculated from BM-J0.2-G100 simulation (dashed line) is an order of magnitude larger than
that from the BM-J0.2-G20 simulation. The off-axis light curves from the BM-J0.2-G100 simulation rise up earlier. At a later time, the afterglow light curves from both simulations overlap with each other.

\subsection{Characteristics of on- and off-axis top-hat BM light curves}
The top-hat jet model can explain the ``jet break'' phenomena observed in GRB afterglows.  On-axis observers will start to see the edge of the jet when its Lorentz factor drops to $\gamma\sim 1/\theta_{\rm{jet}}$. The missing flux will lead to a break in the slope of observed light curves. Before the jet break time, the temporal slope of  high frequency light curves ($\nu>\nu_c$) scales as $F_{\nu}\propto t^{-3p/4+1/2}$. This should be the same for both spherical explosion and top-hat jets. After that, the light curve of the finite top-hat jet scales as $F_{\nu}\propto t^{-p}$ (see e.g. \citealt{1999ApJ...525..737R,1999ApJ...519L..17S}).


    

\begin{deluxetable}{lll}[htb!]  
  \tablecolumns{3}
  \tablewidth{2pc}
  \tablecaption{Synchrotron radiation parameters \label{tab:sync}}
    
  \tablehead{
   \colhead {Variable} & \colhead {BM models}  &
   \colhead {Structured Jet (SJ) models}
  }
  \startdata
  $\epsilon_e$  & $0.1$  & $0.05$  \\
  $\epsilon_B$   & $0.1$ & $0.005$  \\
  $p$            & $2.5$ & $2.2$  \\
  $d_L$          & $2.05 \times 10^{28}$ cm    & $2.05 \times 10^{28}$ cm \\
  $z$            &  $1$            & $1$ \\
  \enddata
  \tablecomments{ Two sets of micro-physical parameter values are utilized for simulations presented in this study.
    Blandford-McKee (BM) models include the BM-J0.2-G20, BM-J0.2-G100, and BM-J0.1-G100 simulation. SJ models include the SJ0.1-EH, SJ0.1-EL, and SJ0.2-EH simulation.
    When we make comparison between SJ and BM models. The BM models use the same set of parameter values with SJ models.}
\end{deluxetable}

\begin{figure}[!ht]
  \centering
  \includegraphics[width=0.45\textwidth]{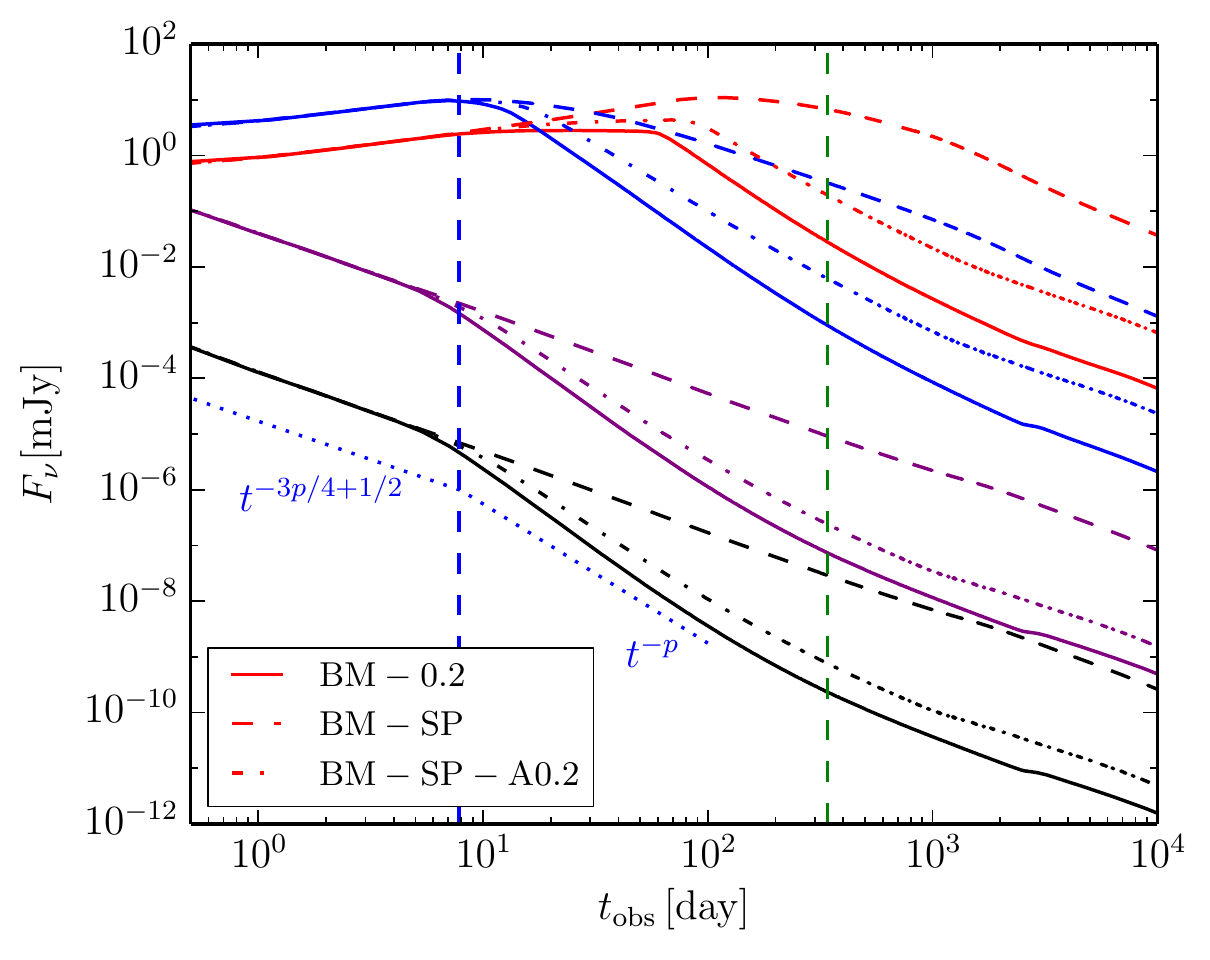}
  \caption{ The on-axis light curves from different BM simulations performed with \texttt{JET}.
    All of the BM simulations start with $t_0=4.37\times 10^6\ s,\rm{and}\,\gamma_0=100$. Solid lines represent light curves calculated from the top-hat BM simulation with jet half opening angle
    0.2. Dashed lines show light curves calculated from a spherical BM
    simulation. Dotted-dashed lines represent the light curves coming from
    a conical segment taken from the spherical BM profile. The half opening angle of the conical segment is $\theta=0.2$.
    From top to bottom, light curves at various frequencies are included -- $10^9$ Hz (red), $10^{11}$ Hz (blue),
    $10^{15}$ Hz (purple), and $10^{17}$ Hz (black). Analytic result of the temporal slope before and after jet break is also shown. Vertical lines indicate the jet break and Newtonian transition time. \label{fig:lc_bm_angle}} 
\end{figure}

                                        
We perform a spherical BM simulation with parameters
identical to the BM-J0.2-G100 model. The light curves from the spherical BM simulation (see dashed lines in Figure \ref{fig:lc_bm_angle}) do not show any break. When we cut a conical segment with half opening angle $\theta=0.2$, and only add emission from this region, the calculated light curves (dotted-dashed line) display the expected temporal break due to pure relativistic beaming. For the light curve of top-hat BM-J0.2-G100 simulation (Solid line), the
jet break happens at around the same time but with a steeper temporal slope. The extra decay then originates from a well-know hydrodynamic effect: lateral spreading of the jet \citep{1999ApJ...525..737R,1999ApJ...519L..17S,2001grba.conf..312G,2009ApJ...698.1261Z,2011ApJ...738L..23W,2012ApJ...751..155V,2012MNRAS.421..570G,2018ApJ...865...94D}. The comparison among these three sets of light curves indicates that the jet break phenomenon due to hydro-dynamical effect is not negligible even for the relatively simplified model considered here. 


\begin{figure}[!ht]
  \centering
  \includegraphics[width=0.45\textwidth]{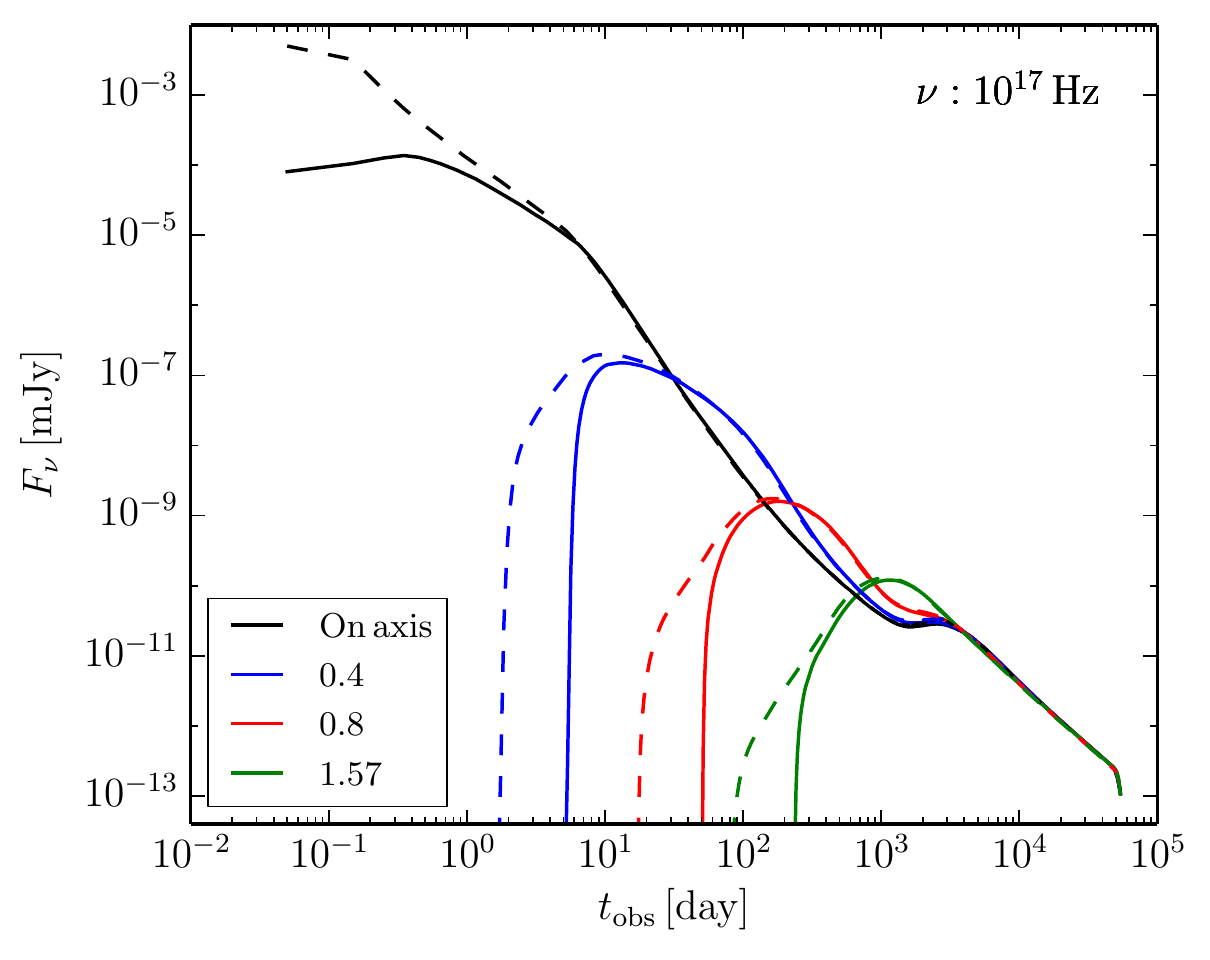}
  \caption{The on- and off-axis X-ray ($10^{17}\,\rm{Hz}$) light curves calculated from two top-hat BM simulations
    performed with \texttt{JET}. The solid lines represent results from the top-hat BM simulation with an initial time 
    $t_0=1.278\times 10^7\ \rm{s}\,(148\ \rm{days})$ and an initial fluid Lorentz factor $\gamma_0=20$. Dashed
    lines represent the top-hat BM simulation with $t_0=4.371\times 10^6\,\rm{s},\gamma_0=100$. \label{fig:lc_g20_g100_offaxis}}
\end{figure}

For off-axis light curves, a natural prediction from top-hat jet models is
the existence of ``orphan'' afterglows. An observer located outside of the
opening angle of relativistic jets will not be able to detect the early
high energy emission due to relativistic beaming. At a later time when the Lorentz factor of the
jets reduces to $\gamma=1/(\theta_{\rm{obs}}-\theta_{\rm{jet}})$, the off-axis observer
 starts to receive emission from the central jet at lower energies. It is then 
possible to detect the afterglow radiation without having detected prompt emission for off-axis observers. As shown in Figure \ref{fig:hk_compare}, the off-axis afterglow emission show features of late rise-up, on a time scale of days to months depending on the viewing angle. However, as we show in Figure \ref{fig:lc_g20_g100_offaxis}, changing the initial Lorentz factor from 20 to 100 leads to an early rise-up for off-axis light curves. The explanation is that the BM-J0.2-G100 profile is set at an earlier time $t_0=4.37\times 10^6\ s$, and at a smaller radius $R_0=1.31\times 10^{17}\, (\rm{cm})$.

In the rest of the paper, we utilize FTD simulations to study the on- and off-axis synchrotron light curves of LGRB jets. We discuss features revealed from  the light curves and their implications. 

\section{Full-Time-Domain Jet Simulation}\label{sec:star}
\subsection{Initial numerical setup }
The stellar progenitor before collapse utilized in the simulations follows the analytical model in \citealt{2015ApJ...806..205D}. This model approximates the output of a MESA \citep{2011ApJS..192....3P,2013ApJS..208....4P} simulation where a low-metallicity rapidly rotating star evolves to a Wolf-Rayet star. The density as a function of radius is:
\begin{equation}\label{eq:density}
\rho(r,0) = \frac{\rho_c(\rm{max}(1-r/R_3,0))^n}{1+(r/R_1)^{k_1}/(1+(r/R_2)^{k_2})}+\rho_{\rm{wind}}(r/R_3)^{-2}.
\end{equation}

The parameters in the above Equation are listed in Table 1 of \cite{2015ApJ...806..205D} and are included here for clarity (see Table
\ref{tab:stellar}). The velocity and pressure are initially set to negligible values. Self-gravity
and stellar rotation are not included in the simulations.

\begin{deluxetable}{llc}[htb!]  
  \tablecolumns{3}
  \tablewidth{2pc}
  \tablecaption{Stellar Parameters\label{tab:stellar}}
  \tablehead{
   \colhead {Variable} & \colhead {Definition}  &
   \colhead {Value} 
  }
  \startdata
  $M_0$ & Characteristic Mass Scale &
  $2\times 10^{33}\rm{g}$\\
  $R_0$ & Characteristic Length Scale & $7\times 10^{10}\rm{cm}$ \\
  $\rho_c$ & Central Density & $3\times 10^7 M_0/R_0^3$\\
  $R_1$ & First Break Radius & 0.0017 $R_0$ \\
  $R_2$ & Second Break Radius & 0.0125$R_0$ \\
  $R_3$ & Outer Radius & 0.65  $R_0$ \\
  $k_1$ & First Break Slope & $3.24$ \\
  $k_2$ & Second Break Slope & $2.57$ \\
  $n$  & Atmosphere Cutoff Slope & $16.7$ \\
  $\rho_{\rm{wind}}$ & Wind Density & $10^{-9}M_0/R_0^3$ \\
  \enddata
  \tablecomments{Courtesy of Table 1 in \cite{2015ApJ...806..205D}.}
\end{deluxetable}

\begin{figure}[!ht]
  \centering
  \includegraphics[width=0.9\columnwidth]{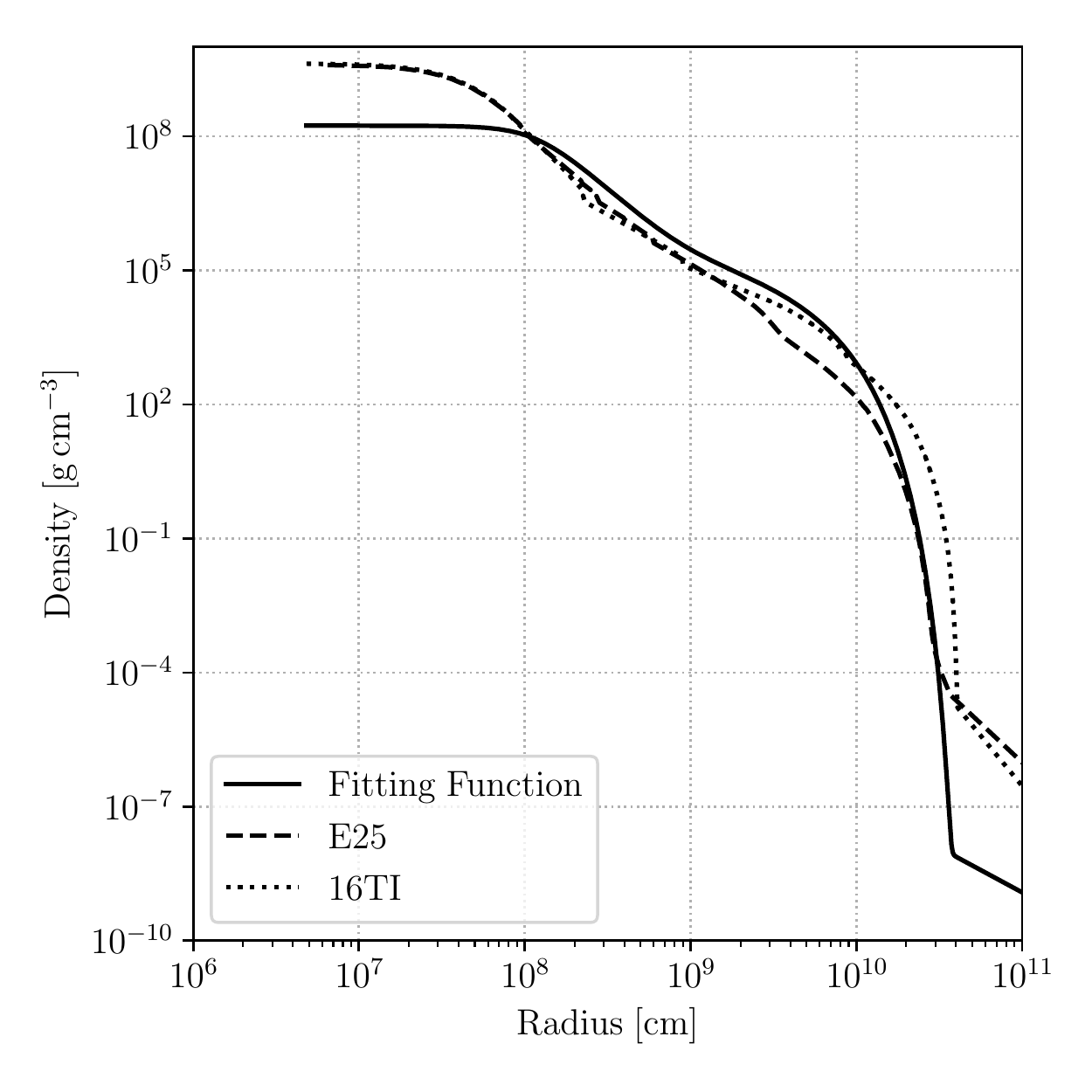}
  \caption{The density profile of the progenitor. A fitting function (Equation \ref{eq:density}) is utilized here following \citealt{2015ApJ...806..205D}. Density profiles from the E25 model in \citealt{2000ApJ...528..368H} and the 16TI model in \citealt{2006ApJ...637..914W} are plotted  for comparison.}\label{fig:density}
\end{figure}
The jet engine is initiated at around the radius $r_0=0.01R_0\ (7\times 10^8\,\rm{cm})$ using a source term. From this distance, the density field of the progenitor in which the jet propagates is comparable to that of the E25 model in \citealt{2000ApJ...528..368H} and the 16TI model in \citealt{2006ApJ...637..914W} (see Figure \ref{fig:density}).
The jet engine model utilizes the  nozzle function  $g(r,\theta)$ in \cite{2015ApJ...806..205D}. For clarity, the expressions are included in the following:
\begin{equation} g(r,\theta) \equiv (r/r_0) e^{-(r/r_0)^2 /2}e^{ (\rm{cos}
\theta - 1) /\theta_0^2 }/N_0, \end{equation}

where $N_0$ is the normalization of $g$ via the integration over $r\in[0,\infty],\theta\in[0,\pi/2]$:

\begin{equation} N_0 \equiv 4 \pi r_0^3 ( 1. - e^{-1/\theta_0^2} )
\theta_0^2. \end{equation}

The source terms in Equations (\ref{eq:hydro1}) and (\ref{eq:hydro2})
are given in the following:

\begin{eqnarray}
S^0 & = & L_0 e^{-t/\tau_0} g(r,\theta), \\
S^r & = & S^0 \sqrt{ 1 - 1/\gamma_0^2 }, \\ 
S_D & = & S^0/\eta_0. 
\end{eqnarray}

This jet engine features a smoothly decaying tail with an average engine duration $\tau_0=10\,s$. The engine completely shuts down at around $20\,s$ (see discussions of engine duration in e.g. \citealt{2013MNRAS.436.1867L}). 
The simulation (denoted as SJ0.1-EH (Energy High) hereafter) performed in this study differs from \citealt{2015ApJ...806..205D} in two ways.
First, we utilize the incorporated AMR scheme which enforces the criteria $\Delta r/r < 1/(16\Gamma^2)$ to better resolve the relativistic shell. Second, we adopt the newly implemented RC-EOS instead of the original ideal gas EOS. We also perform additional simulations. One with relatively low jet engine energy (denoted as SJ0.1-EL). Another with
a different jet engine half opening angle  $\theta_0=0.2$ (denoted as SJ0.2-EH). The jet engine parameter values for these three models are listed in Table \ref{tab:engine}.

\begin{deluxetable}{lccc}[htb!]  
  \tablecolumns{4}
  \tablewidth{2pc}
  \tablecaption{Jet Engine Parameters\label{tab:engine}}
  \tablehead{
   \colhead {Variable} & \colhead {SJ0.1-EH}  &
   \colhead {SJ0.1-EL} & \colhead {SJ0.2-EH}
  }
  \startdata
  $L_{0}\,\rm{[erg\,s^{-1}]}$ & $1.5\times 10^{51}$ &
  $1.5\times 10^{50}$ & $1.5\times 10^{51}$ \\
  $\tau_{0}\,\rm{[s]}$ & $10$  & $10$ & $10$\\
  $\eta_0$ & 100  & 100 & 100\\
  $\gamma_0$ & 50 & 50 & 50\\
  $\theta_{\rm{0}}$ & $0.1$ & $0.1$ & $0.2$ \\
  \enddata

  \tablecomments{ $L_{0}$ and $\tau_0$ represents jet engine power and average jet engine duration, respectively. The Energy-to-Mass Ratio $\eta_0$ and injected Lorentz factor $\gamma_0$ is set to the same value for all of the jet engine models. $\theta_{\rm{0}}$ is the half opening angle of the injected jet engines.}
\end{deluxetable}

To better interpret the role of each fluid component, we use three passive scalars $X_i$ to track the mass fraction of each individual component filling the cells. The subscript $i$ labels each individual component: 1 for stellar progenitor, 2 for ISM material, and 3 for jet engine material.
  Initially, $X_1\,(X_2)$ is set to $1\,(0)$ inside the progenitor and $0\,(1)$ in the ISM. Three auxiliary equations are solved accordingly \citep{2013ApJ...775...87D}:
  \begin{equation}
    \partial_\mu (X_i \rho u^\mu) = S_D\, . \label{eq:hydro_passive}\\
  \end{equation}
The conserved mass for each component ($M_i=\int X_i\rho u^0 dV$) is updated based on the density flux through the cell boundary and the addition of source term (for the injection of jet engine material). Dividing the individual conserved mass $M_i$ by the total conserved mass ($M=\int \rho u^0 dV$) gives us the new primitive passive scalar $X_i$.

\subsection{Dynamical details of the full-time-domain jet simulation}

\begin{figure*}[!ht]
  \centering  
  \subfloat[\label{fig:j0.1_EH_4panel_1s}]{
    \includegraphics[clip,width=1\columnwidth]
                        {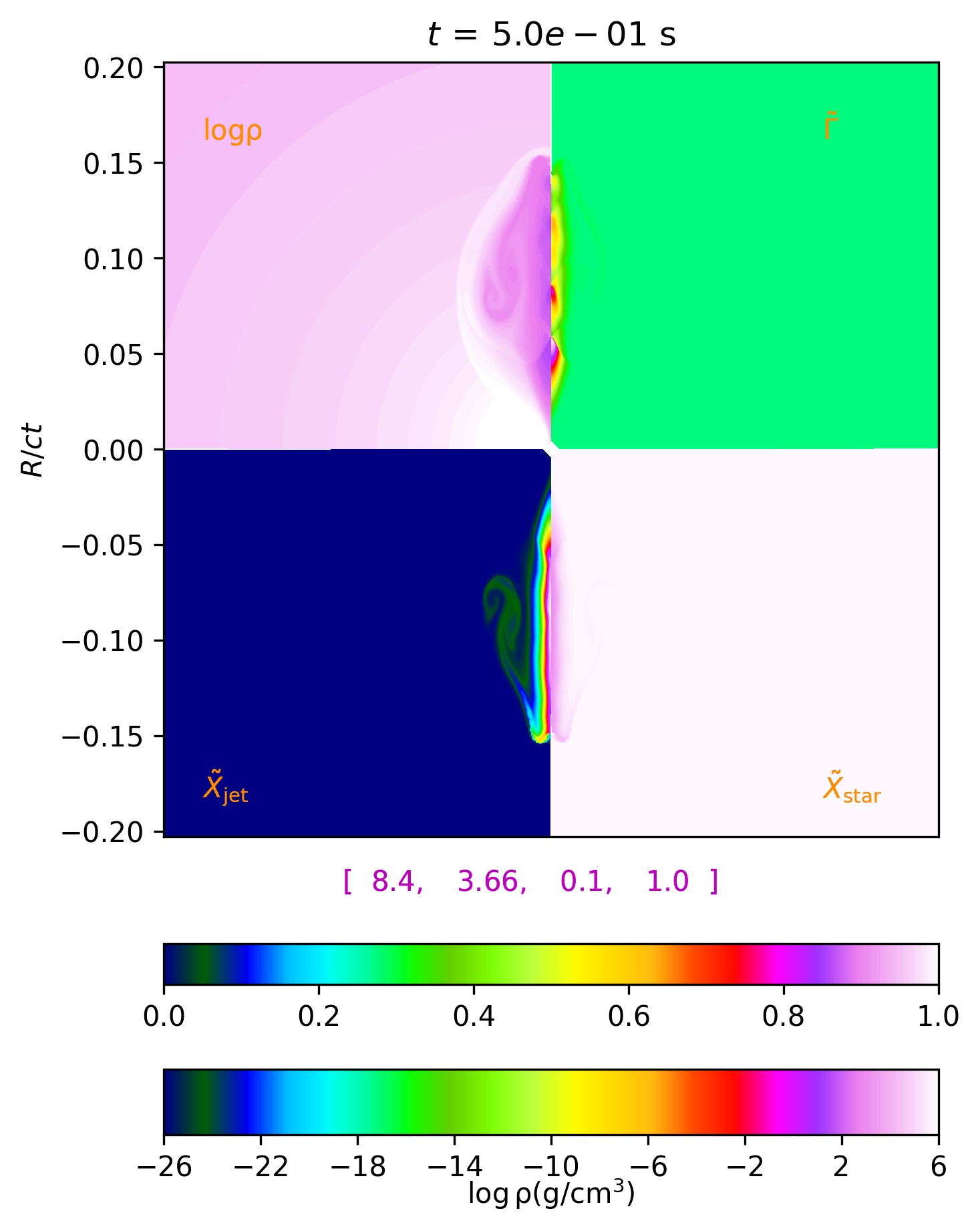}
  }
  \subfloat[\label{fig:j0.1_EH_4panel_2s}]{
    \includegraphics[clip,width=1\columnwidth]
                        {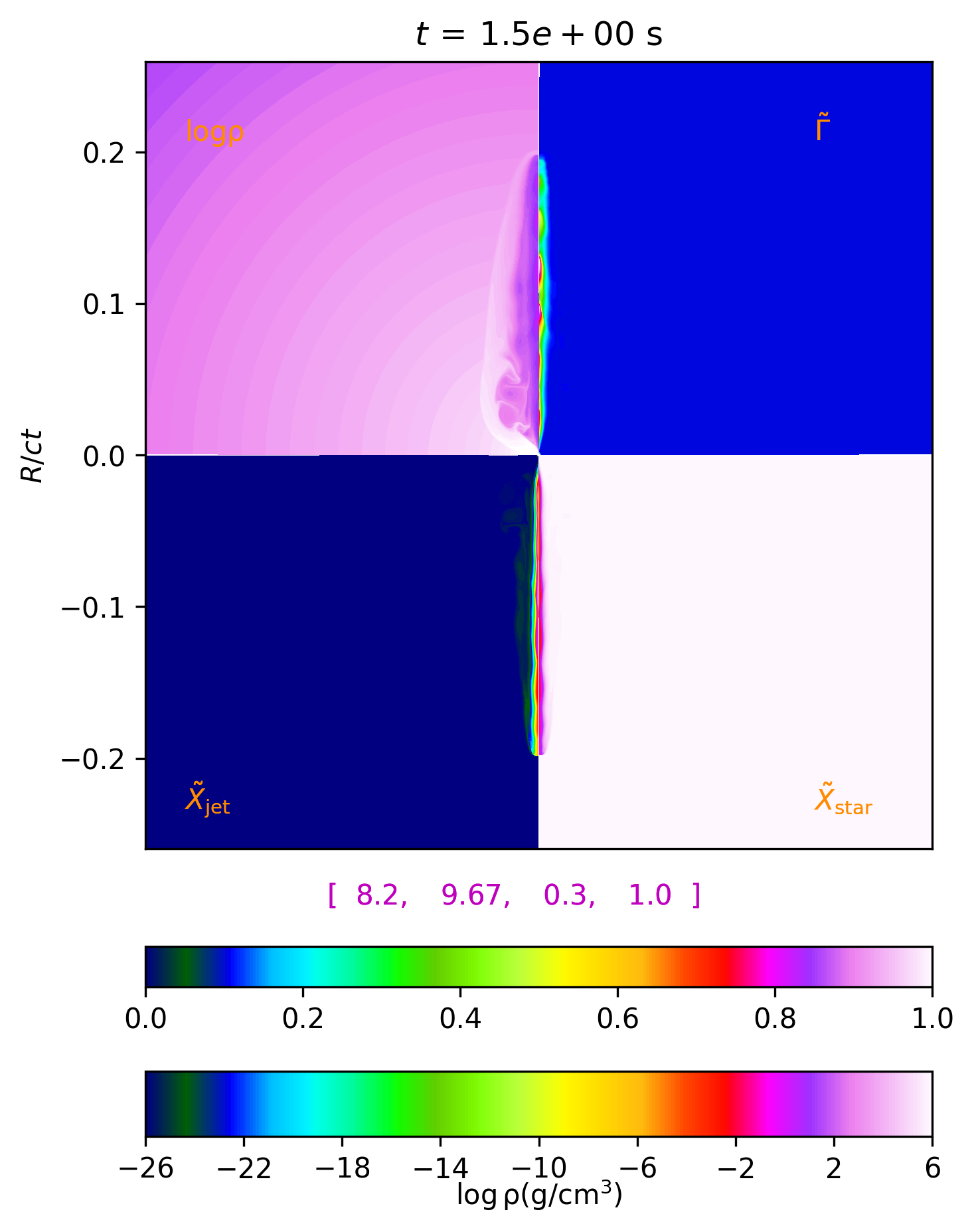}
  }
  \caption{Early time snapshots of jet propagation from the SJ0.1-EH simulation. Each
    snapshot displays four panels: the upper left and upper right panel shows the contour plot of logarithmic density $\rm{log}\,\rho$ and  normalized Lorentz factor $\hat{\Gamma}$, respectively. The contour plots of normalized mass fraction of jet engine material $\hat{X}_{\rm{jet}}$ and stellar-mass material $\hat{X}_{\rm{star}}$ are shown in the lower left and lower right panel. The values in the square bracket, represent the maximum value of [log density, Lorentz factor, jet-engine material fraction, stellar material fraction] in the simulation domain. These values are also the ones used in the normalization of contour plots. The lab frame time $t$ is shown in the title of each snapshot. The length of the simulation domain is scaled by $ct$.   \label{fig:4panel_early}}
\end{figure*}

\begin{figure*}
  \centering
  \includegraphics[width=0.8\textwidth]{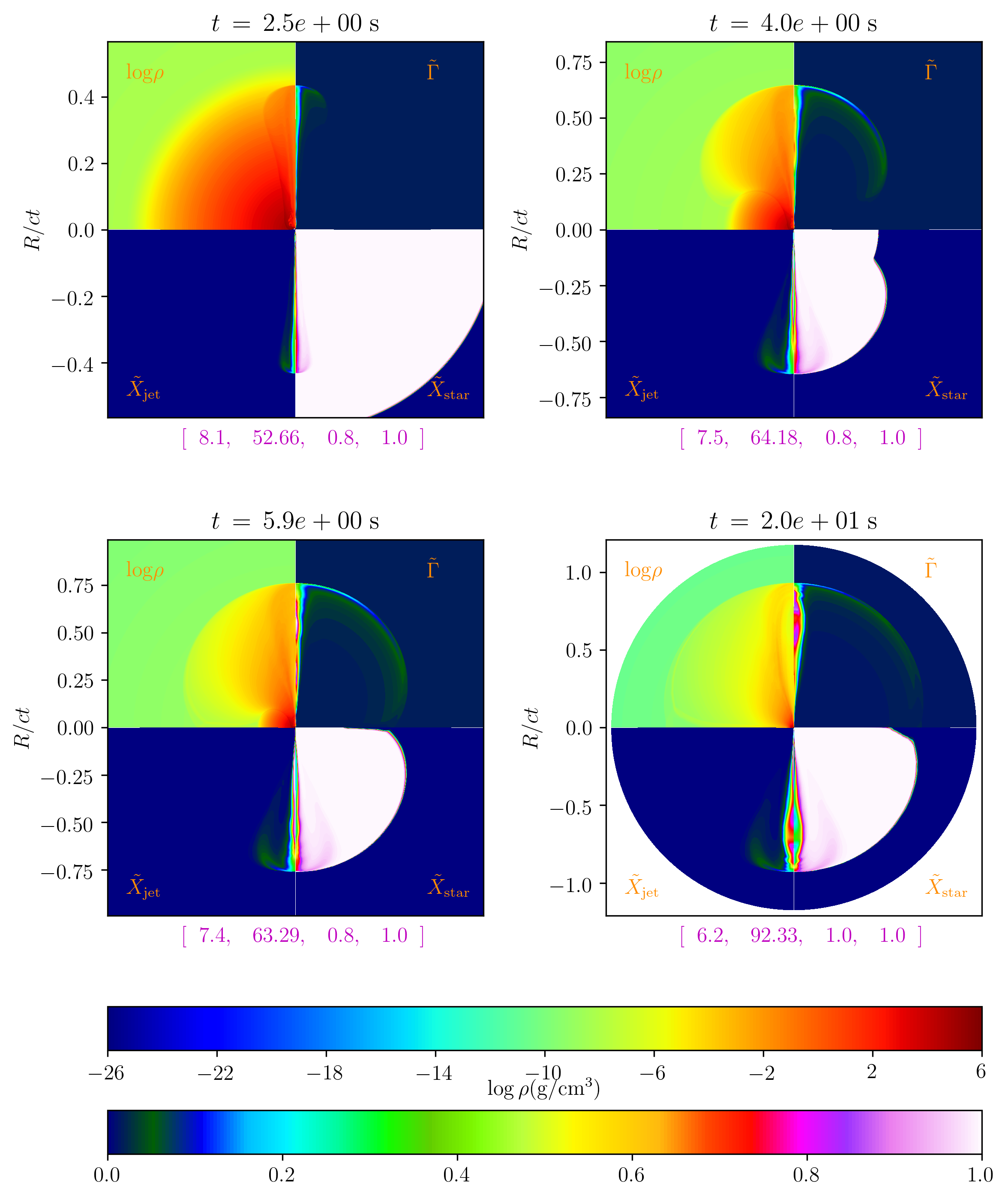}
  \caption{Early time snapshots of physical variables from the SJ0.1-EH simulation. Each
    snapshot displays four panels: the upper left and upper right panel shows the contour plot of logarithmic density $\rm{log}\,\rho$ and  normalized Lorentz factor $\hat{\Gamma}$, respectively.
    The contour plots of normalized mass fraction of jet engine material $\hat{X}_{\rm{jet}}$ and stellar-mass material $\hat{X}_{\rm{star}}$ are shown in the lower left and lower right panel.
    The values in the square bracket, represent the maximum value of [log density, Lorentz factor, jet-engine material fraction, stellar material fraction] in the simulation domain. These values are also the ones used in the normalization of contour plots. The
    lab frame time $t$ is shown in the title of each snapshot. 
    The length of the simulation domain is scaled by $ct$. \label{fig:dynamics_A}}.
\end{figure*}  

\begin{figure*}
  \centering
  \includegraphics[width=0.8\textwidth]{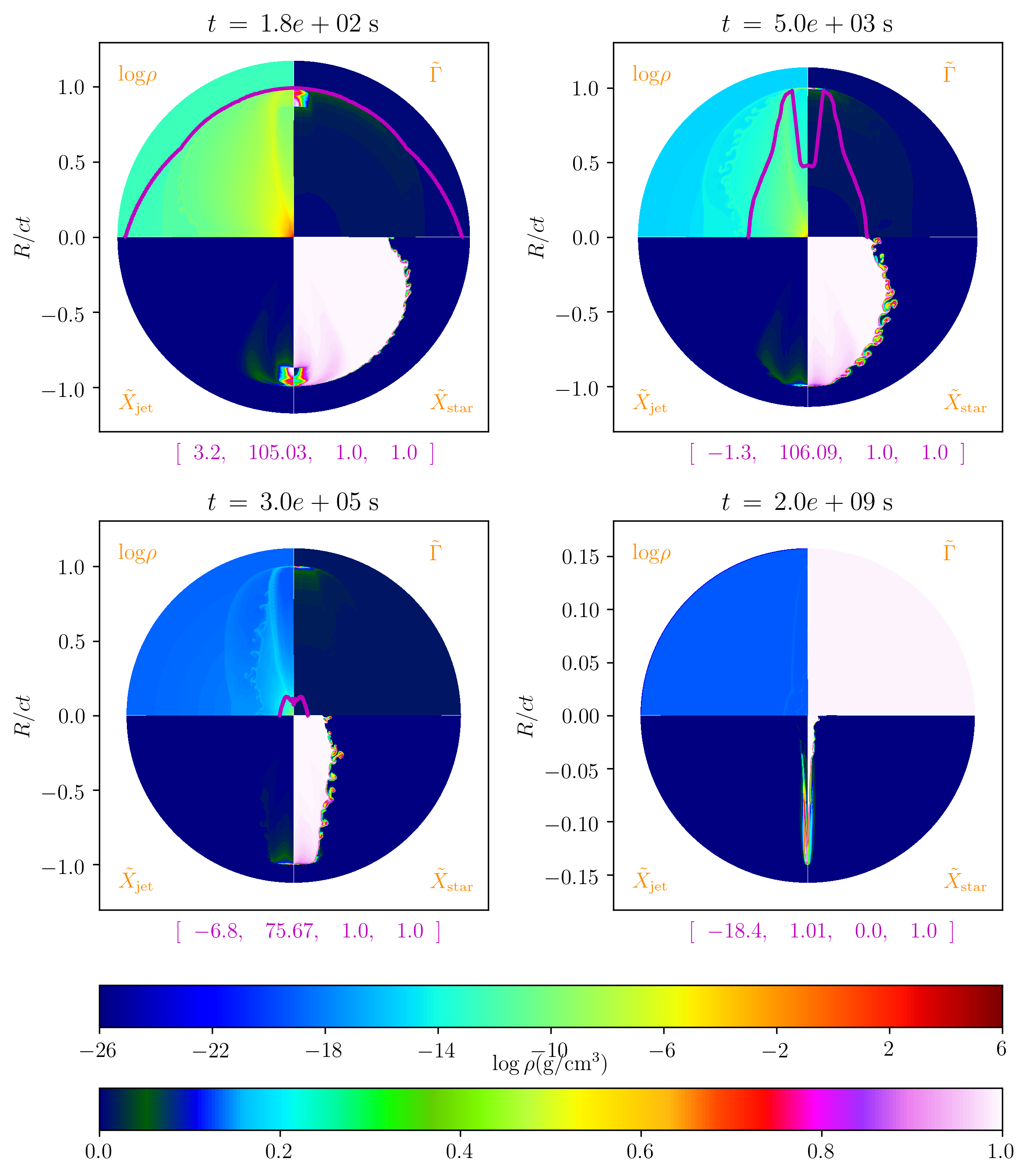}
  \caption{Late time snapshots of physical variables from the SJ0.1-EH simulation. Each
    snapshot displays four panels: the upper left and upper right panel shows the contour plot of logarithmic density $\rm{log}
    \,\rho$ and  normalized Lorentz factor $\hat{\Gamma}$, respectively.
    The contour plots of normalized mass fraction of jet engine material $\hat{X}_{\rm{jet}}$ and stellar-mass material $\hat{X}_{\rm{star}}$ are shown in the lower left and lower right panel.
    The values in the square bracket, represent the maximum value of [log density, Lorentz
    factor, jet-engine material fraction, stellar material fraction] in the simulation domain. These values are also the ones used in the normalization of contour plots. The
    lab frame time $t$ is shown in the title of each snapshot. 
    The length of the simulation domain is scaled by $ct$. The magenta line represents
    the photosphere location viewed by on-axis observers. \label{fig:dynamics_B}}.
\end{figure*}  
\subsubsection{launch of jet engine}
Figure \ref{fig:4panel_early} shows the early evolution of jet engine flow upon injection. The ram pressure generates a bow shock. Dense stellar material gets pushed to the side, forming a cocoon which confines the engine outflow. A high-density wedge of stellar material develops at the head of the jet. It shreds jet engine materials from the head. These engine materials curl back, forming vortexes. These vortexes then detach from the jet and get swept backward relative to jet propagation. This vortex shedding phenomena is, generally speaking, similar to that found in previous simulations \citep{2002MNRAS.331..615S,2004ApJ...606..804M,2007ApJ...665..569M}. The speed of jet head keeps increasing, soon exceeds the local sound speed (At the time $t=1.5s$, the Lorentz factor of the head of jet is $\sim 2$ as shown in Figure \ref{fig:4panel_early}). The backflow becomes quasi-straight to the main jet \citep{2010ApJ...709L..83M}. At early times, the bow shock has a narrow head and a wide tail. When it approaches the surface, the jet head expands in the low density envelope as shown in the next subsection (see also e.g. \citealt{2003ApJ...586..356Z,2004ApJ...608..365Z}). We use spherical coordinate to conduct simulations. As jets propagate outward, the width of the cell increases. Features of the inner part of jet (close to pole) may not be fully resolved. We well resolve the relativistic shell in the radial direction via previously described AMR scheme. As the ultra-relativistic jet shell penetrates the progenitor, the stellar material that lies on top of the jet easily gets pushed aside. No strong ``plug'' instability has been seen \citep{2010ApJ...717..239L,2013ApJ...777..162M,2018MNRAS.473..576G,2018ApJ...863...58X}.
  
\subsubsection{propagation of the jet}
Figures \ref{fig:dynamics_A} and \ref{fig:dynamics_B} show snapshots of the simulation domain at different stages of jet evolution for the SJ0.1-EH model. Each snapshot displays the contour of log density, normalized Lorentz factor, and normalized mass fraction of jet engine/stellar material. At lab frame time $t=2.5\,\rm{s}$, the jet outflow begins to expand in the low-density wind. The shock wraps around the star as shown in the snapshots at $t=4\,\rm{s}$ and $t\approx 6\,\rm{s}$. At a later time, $t=20\,\rm{s}$, the shock accelerates and approaches its terminal Lorentz factor $\sim 10^2$. The engine completely turns off at around this time. Along the polar axis, a relativistic blob forms behind the shock front. Through internal collisions,  the relativistic blob forms an ultra-relativistic thin shell. At first, the relativistic shell is hidden behind the photosphere (indicated by the magenta line in Figure \ref{fig:dynamics_B}). We define the photosphere as the place where the optical depth is unity. We estimate the optical depth according to:
\begin{equation}
  \tau = \int^{\infty}_{r_{\rm{ph}}}\sigma_T \Gamma(1-\beta \rm{cos\theta})n dl\,,\label{eq:optical}
\end{equation}
where $\sigma_T$ is the Thomson scattering cross section, $\beta$ is the absolute value of the velocity normalized by speed of light, $\Gamma$ is the Lorentz factor of the gas,  $\theta$ is the angle
between the velocity vector and the line of sight, $n$ is the electron number density \citep{1991ApJ...369..175A,2011ApJ...732...26M}. In detail, we first initiate sufficient number of tracing rays at the given observing angle. Each tracing ray will enter the spherical domain from a position of the outer boundary. We then calculate the crossed length in each cell the ray intercepts and perform the integration of optical depth (Equation \ref{eq:optical}), assuming the density is uniform within the cell. The contribution of optical depth from materials outside of the simulation domain is added using an analytical expression. Following this procedure, we're able to get the optical depth for all of the cells  in the simulation domain at each snapshot. Note that, this procedure is a simplified version in terms of the calculation of actual photosphere. Photons are propagating in a turbulent, evolving density background. Integration of optical depth over continous snapshots is then preferred. In this work, we focus on the study of optically-thin synchrotron radiation light curves. Once the emitting shell breaks out of the photosphere, the photosphere position has no impact on the shape of the light curve. We find the characteristics of synchrotron light curves are not sensitive to the exact definition of photosphere. For detailed treatment of photons breaking out of the photosphere, we refer readers to the study of photospheric emission (e.g. \citealt{2011ApJ...732...26M,2011ApJ...732...34L,2018MNRAS.478.4553D,2018MNRAS.479..588G}).

A wind profile is adopted here to describe the density field of surrounding interstellar medium (ISM): $\rho_{\rm{ISM}} = A r^{-2}, A=24\times (5\times 10^{11} \rm{g\,cm^{-1}})$. The photosphere is initially located at a radius  $r_{\rm{ph}} = 4.8 \times 10^{12}\,\rm{cm}$. At around  $\sim 2\times 10^2\,\rm{s}$, the shock front breaks through the original photosphere. The photosphere then advances outward with the jet. Eventually, at around $t \sim 5\times
 10^3\,\rm{s},\,r\sim 10^{14}\,\rm{cm}$, the photosphere begins to fall behind the relativistic shell.
The process of jet breaking out of the photosphere covers the dynamical distance where prompt emission is estimated to occur (e.g. \citealt{1999PhR...314..575P,2015PhR...561....1K}). At $t=3\times 10^5\,\rm{s}$, the photosphere falls far behind the relativistic shell, and Kelvin-Helmholtz instability is seen behind the shock front. At $t\sim 2\times10^9\,\rm{s}$, the jet reaches a distance $\sim 3\,\rm{parsecs}$. At this time, the jet and stellar material forms a highly aspherical structure and has become fully Newtonian. 

\begin{figure*}[!ht]
  \centering  
  \subfloat[\label{fig:j0.1_Eh_structure}]{
    \includegraphics[clip,width=0.3\textwidth]
                        {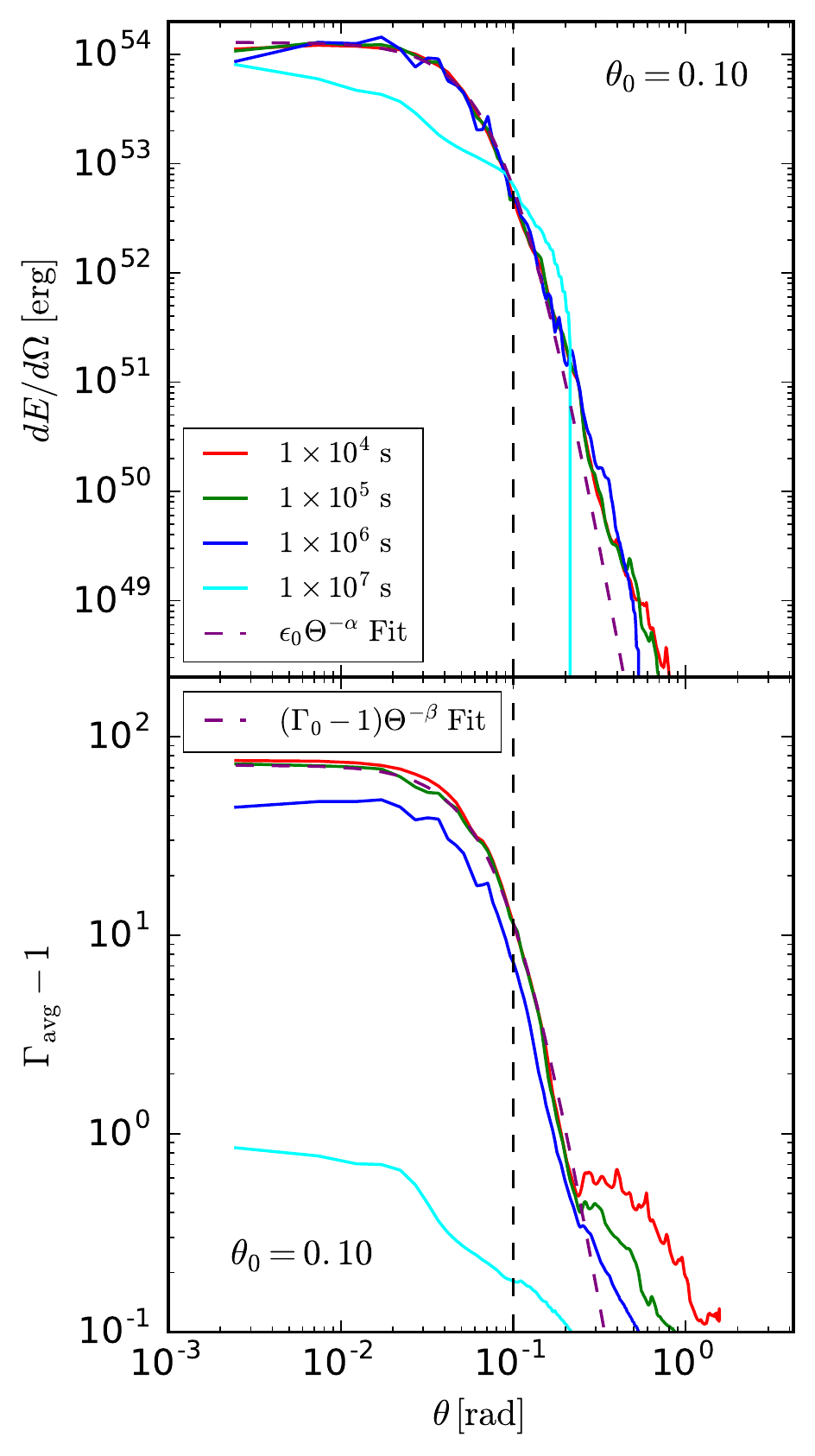}
  }
  \subfloat[\label{fig:j0.1_El_structure}]{
    \includegraphics[clip,width=0.3\textwidth]
                        {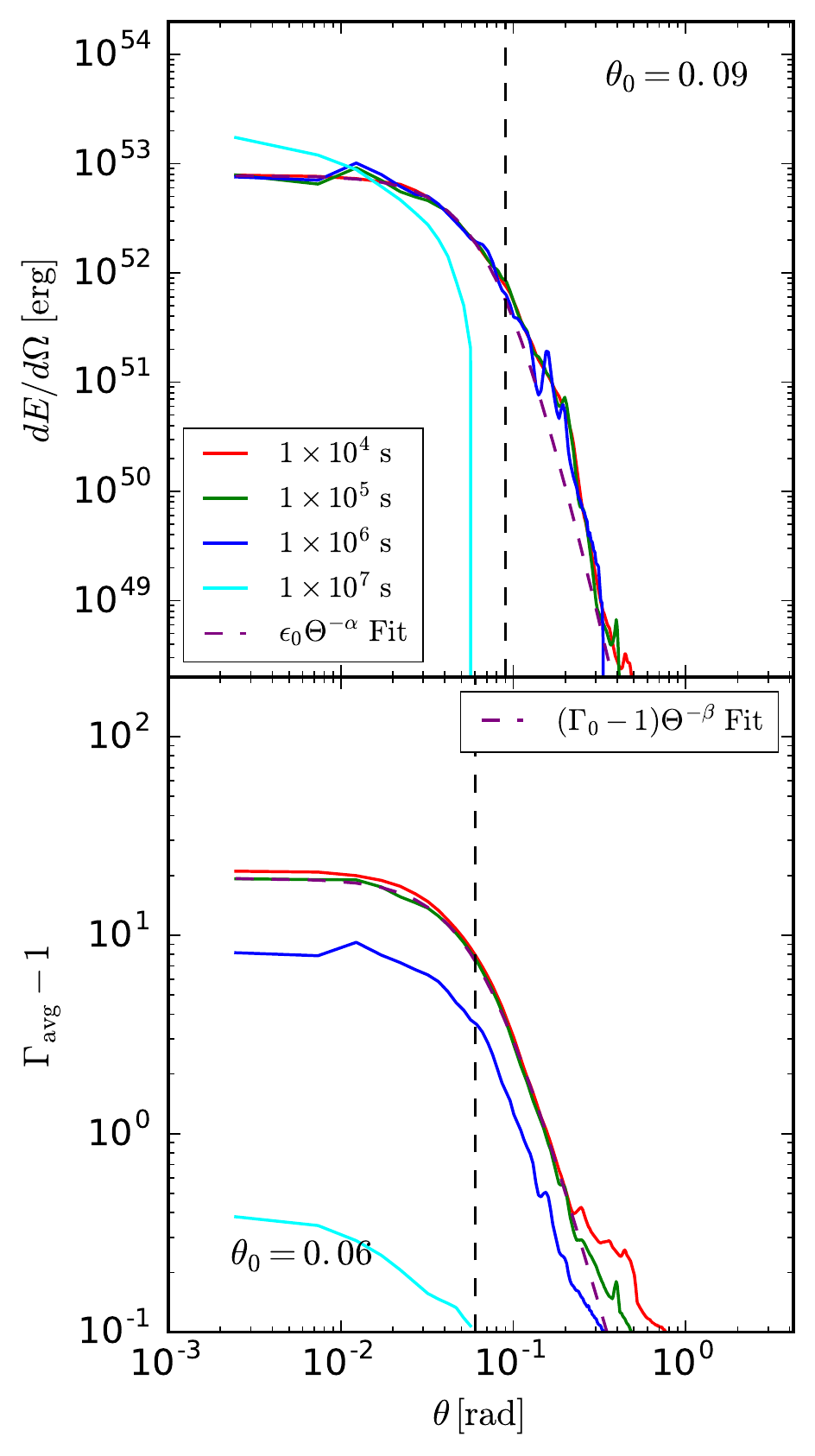}
  }
  \subfloat[\label{fig:j0.2_Eh_structure}]{
    \includegraphics[clip,width=0.3\textwidth]
                        {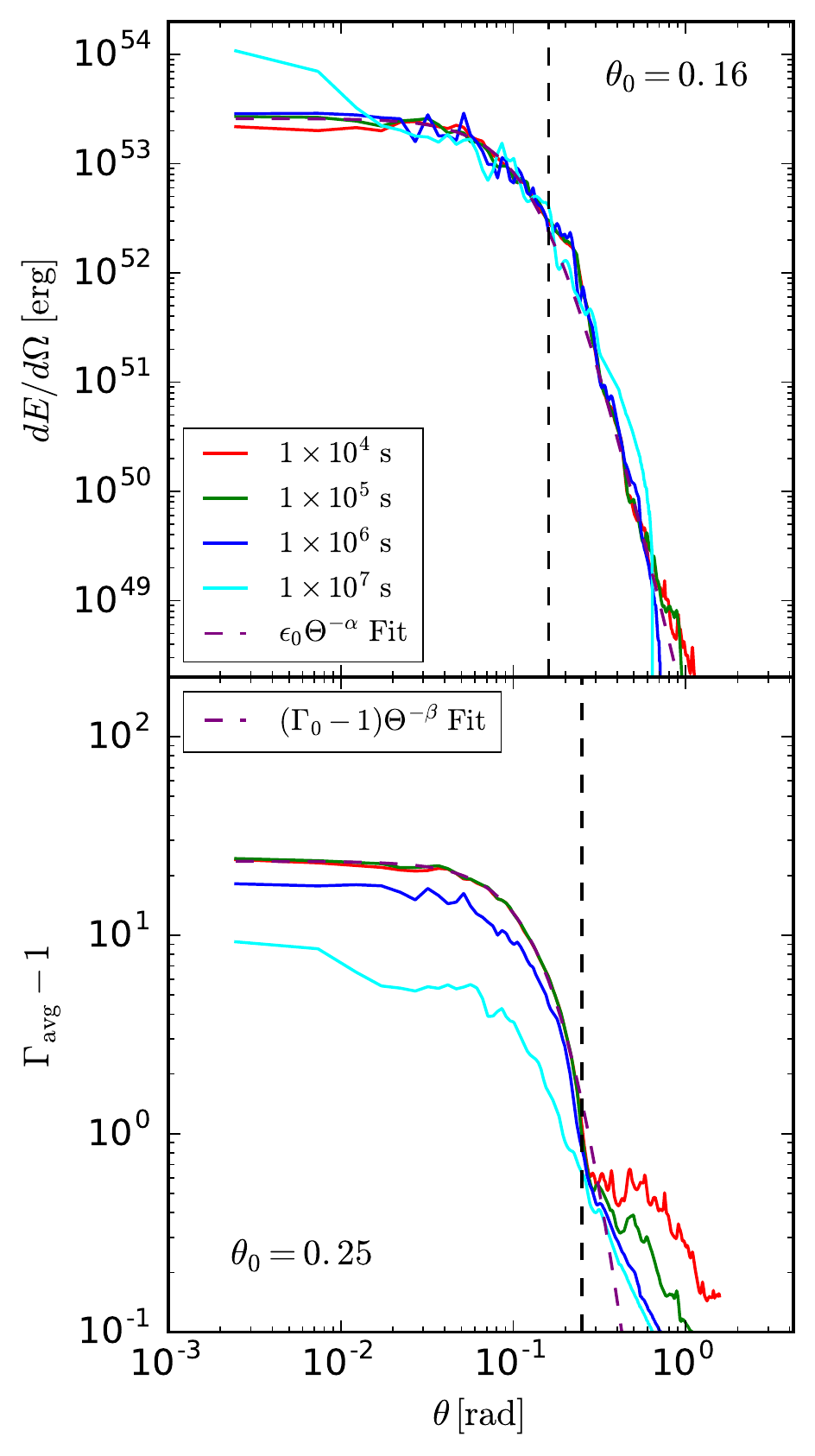}
  }
  \caption{The angular distribution plots of the total energy (top panel) and energy-averaged Lorentz factor (bottom panel) for relativistic shells in SJ0.1-EH simulation (left column), SJ0.1-EL simulation (middle column) and SJ0.2-EH simulation (right column).
    The fitting curve is shown in dashed line for each plot. Equation \ref{eq:angular_energy} and \ref{eq:angular_lorentz} are used to fit the angular distribution of energy 
    and energy-averaged Lorentz factor, respectively \citep{2003ApJ...591.1086G}. The fitting values for these three simulations are listed in Table \ref{tab:fit}. The value of $\theta_0$ for each fitting is denoted by the vertical line in each plot.
    \label{fig:structure}}
\end{figure*}

\begin{deluxetable}{lccc}[htb!]  
  \tablecolumns{4}
  \tablewidth{2pc}
  \tablecaption{Fitting parameter values for the angular structure of emerged jets.\label{tab:fit}}
    
  \tablehead{
   \colhead {Variable} & \colhead {SJ0.1-EH}  &
   \colhead { SJ0.1-EL} & \colhead { SJ0.2-EH }
  }
  \startdata
  $\theta_{\rm{core}}$ &  0.1  & 0.09 & 0.16\\
  $\epsilon_0$ & $\unit[1.3\times 10^{54}]{erg}$ & $\unit[7.8\times 10^{52}]{erg}$ & $\unit[2.6\times 10^{53}]{erg}$ \\
  $\alpha$ & 8.9  & 7.0 & 6.7 \\
  $\Gamma_0$ & 73 & 20 & 25 \\
  $\beta$  &  5.1 & 3.1  & 7.9\\
  \enddata
  \tablecomments{ $\theta_{\rm{core}}$ is defined as the half opening angle of the central core for structured jets. $\epsilon_0,\,\alpha,\,\Gamma_0,\,\beta$ are fitting parameters from Equation \ref{eq:angle} - Equation \ref{eq:angular_lorentz}. }
\end{deluxetable}

\subsection{Angular structure of the jets}
The angular structure of the jets features an ultra-relativistic core, primarily composed of jet-engine material. The core is surrounded by a mildly relativistic sheath (see Figure \ref{fig:dynamics_B}). The angular distribution of the total energy (excluding rest mass energy) $dE/d\Omega$, and energy-averaged Lorentz
factor $\Gamma$ for the emerged jet, are shown in Figure \ref{fig:structure}. During the coasting period $\sim 10^2 - 10^6$ s, the jet angular structure does not change significantly. They can be well fit by a universal structured jet (USJ) model in which 
$dE/d\Omega$ and $\Gamma$ varies as a power law of polar angle \citep{2003ApJ...591.1075K,2003ApJ...591.1086G,2014PASA...31....8G}, 
\begin{eqnarray}
  \Theta &=&\sqrt{1+(\theta/\theta_0)^2}\,,\label{eq:angle}\\
  \epsilon(\theta) &=&\epsilon_0 \Theta^{-\alpha}\label{eq:structure}\,,\label{eq:angular_energy}\\
  \Gamma(\theta) &=& 1 + (\Gamma_0-1)\Theta^{-\beta}.\label{eq:angular_lorentz}
\end{eqnarray}
where $\theta_0,\,\epsilon_0,\,\alpha,\,\beta,\ \rm{and\ }\Gamma_0$ are free fitting parameters. We fit the three performed simulations: SJ0.1-EH, SJ0.1-EL, and SJ0.2-EH, and list their fitting parameter values in Table \ref{tab:fit}. As shown in Figure \ref{fig:structure}, the half opening angle of the central core is $\theta_{\rm{core}}\sim 0.1$ for the SJ0.1-EH and SJ0.1-EL simulations, and  $\theta_{\rm{core}}\sim 0.16$ for the SJ0.2-EH simulation. The total energy of the jet core: $E_{\rm{jet}}\sim 10^{52}\,\rm{erg}$ for SJ0.1-EH and SJ0.2-EH, and $E_{\rm{jet}}\sim 10^{51}\,\rm{erg}$ for SJ0.1-EL, falls within the inferred range of GRB kinetic energy (see e.g. \citealt{2001ApJ...562L..55F,2002ApJ...571..779P,2003ApJ...590..379B,2004ApJ...613..477L,2016ApJ...818...18G}). The angular power-law decay index $\alpha$ in all of the simulations is significantly larger than $2$, the typical value adopted in the USJ model \citep{2002MNRAS.332..945R,2002ApJ...571..876Z,2014PASA...31....8G}. An observational correlation between the isotropic emitted energy $E_{\rm{iso}}$ and the spectral peak energy $E_p$: $E_{\rm{iso}}\propto E_p^2$ has been discovered \citep{2002A&A...390...81A}. This relationship extends from gamma-ray bursts (GRBs) to X-ray flashes (XRFs) (see e.g. \citealt{2005ApJ...620..355L}). In the unified GRB-XRF model, XRFs are the result of a highly collimated GRB jet viewed off axis \citep{2002ApJ...571L..31Y,2005ApJ...620..355L}. The observed $E_p\,(E_{\rm{iso}})$ range is more than $2\,(4)$ orders of magnitude (see e.g. \citealt{2004ApJ...601L.119Z}). For the USJ model with $\alpha=2$, the relation $E_p\propto E_{\rm{iso}}^{1/2}\propto\theta^{-1}$ implies that the viewing angles of XRFs need to be at least 2 orders of magnitude larger than those of GRBs.  This puts strong constraint on the USJ model with $\alpha=2$ (or vice-versa the unified GRB-XRF model). \citealt{2004ApJ...601L.119Z} shows that a quasi-universal Gaussian-like structured jet model with a steeper angular profile can reconcile this. The structured jets emerged in our simulations feature a steep angular profile with $\alpha \sim 8$. This profile can also explain the wide range of observed values for $E_{\rm{iso}}$ given limited off-axis observer angles. We only need to have a viewing angle several times larger than $\theta_{\rm{core}}$ to get an XRF whose $E_{\rm{iso}}$ is $10^2-10^4$ times lower than the typical GRB $E_{\rm{iso}}$ \citep{2004ApJ...601L.119Z}.


\section{Synchrotron light curves from full-time-domain jet simulations}\label{sec:star_rad}

\begin{figure*}[!ht]
  \centering  
  \subfloat[\label{fig:LC_SJ0.1-EH}]{
    \includegraphics[clip,width=1\columnwidth]
                        {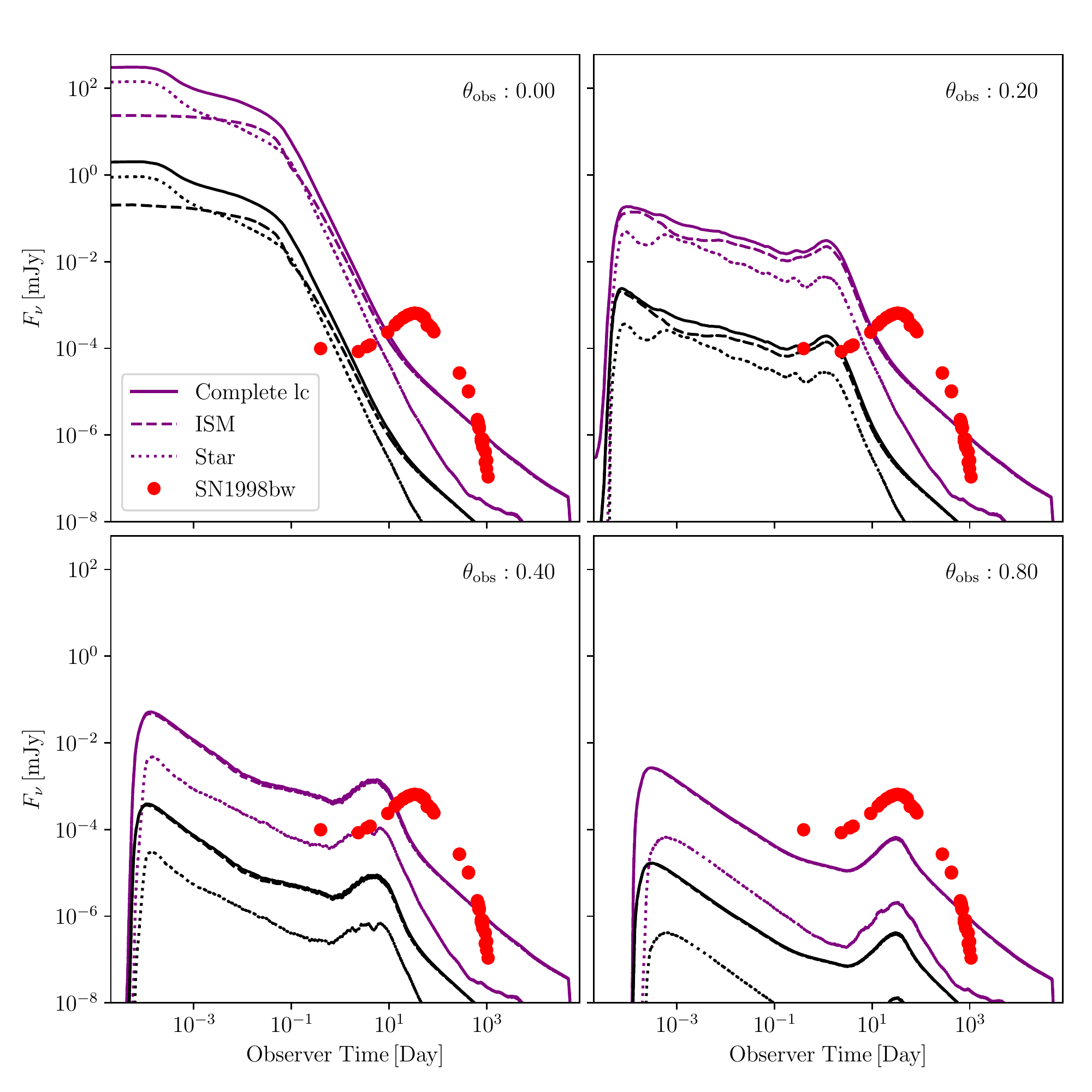}
  }
  \subfloat[\label{fig:LC_SJ0.1-EL}]{
    \includegraphics[clip,width=1\columnwidth]
                        {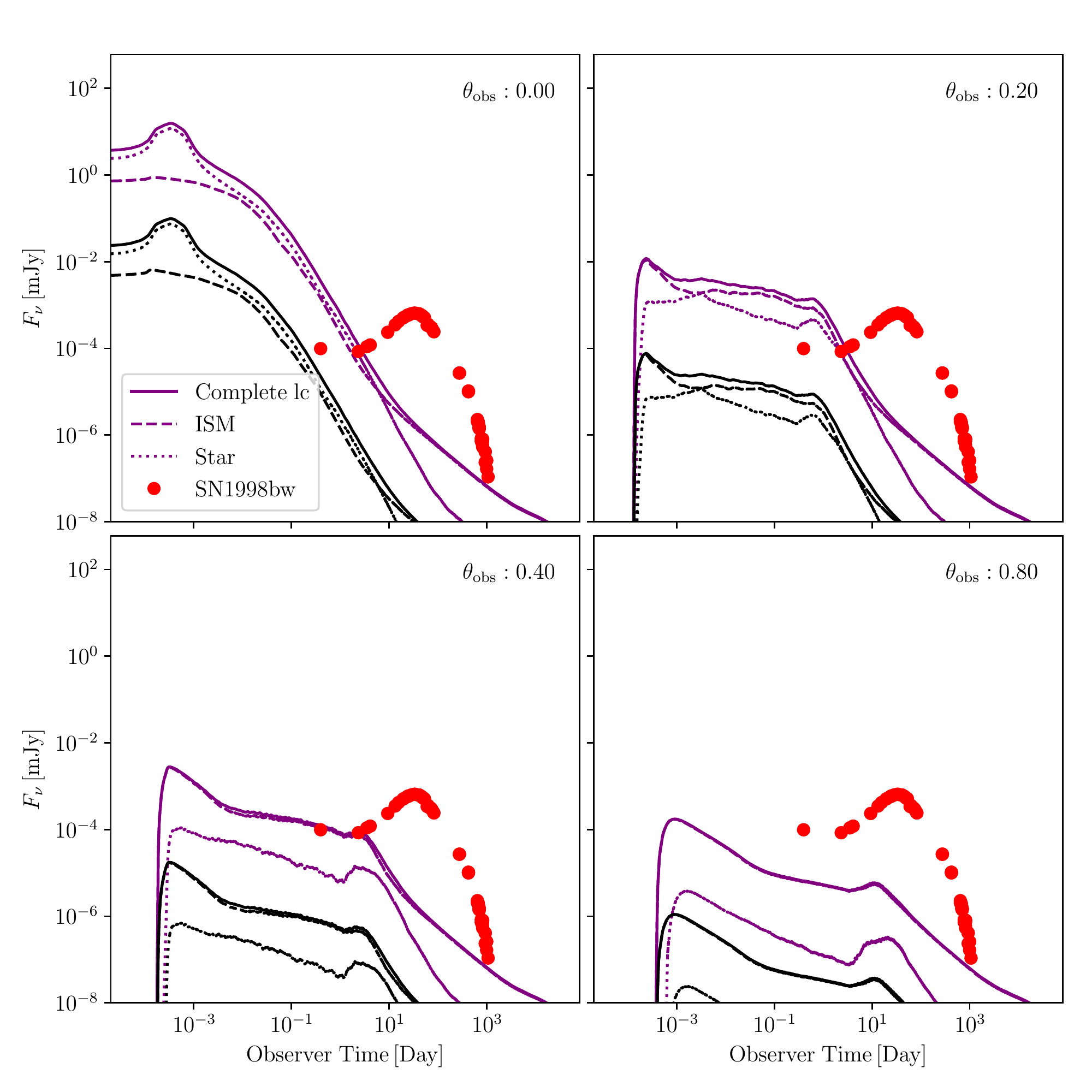}
  }\\[-0.5ex]
  \subfloat[\label{fig:LC_SJ0.2-EH}]{
    \includegraphics[clip,width=1\columnwidth]
                    {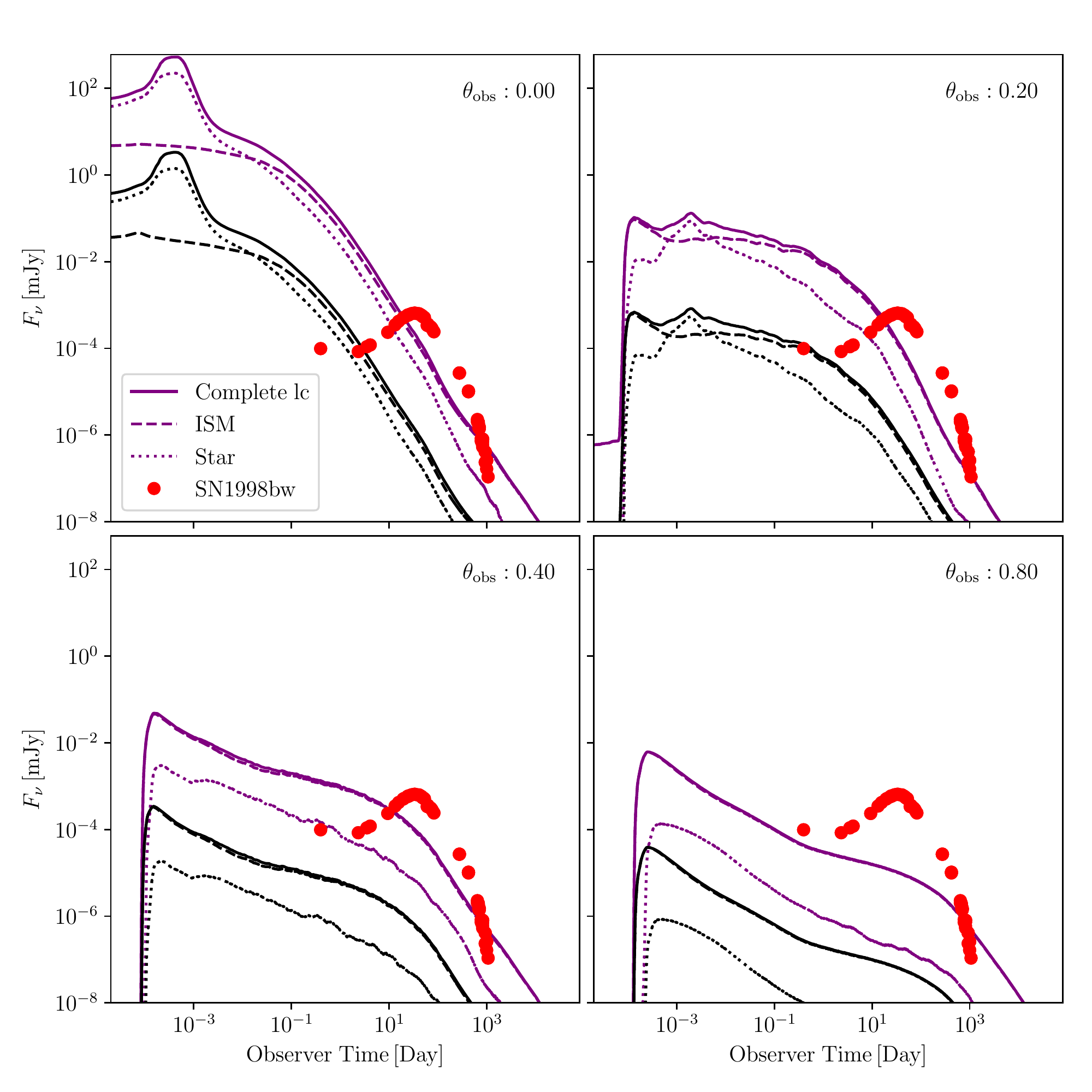}
  }
  \subfloat[\label{fig:LC_grb_bg}]{
    \includegraphics[clip,width=1\columnwidth]
                    {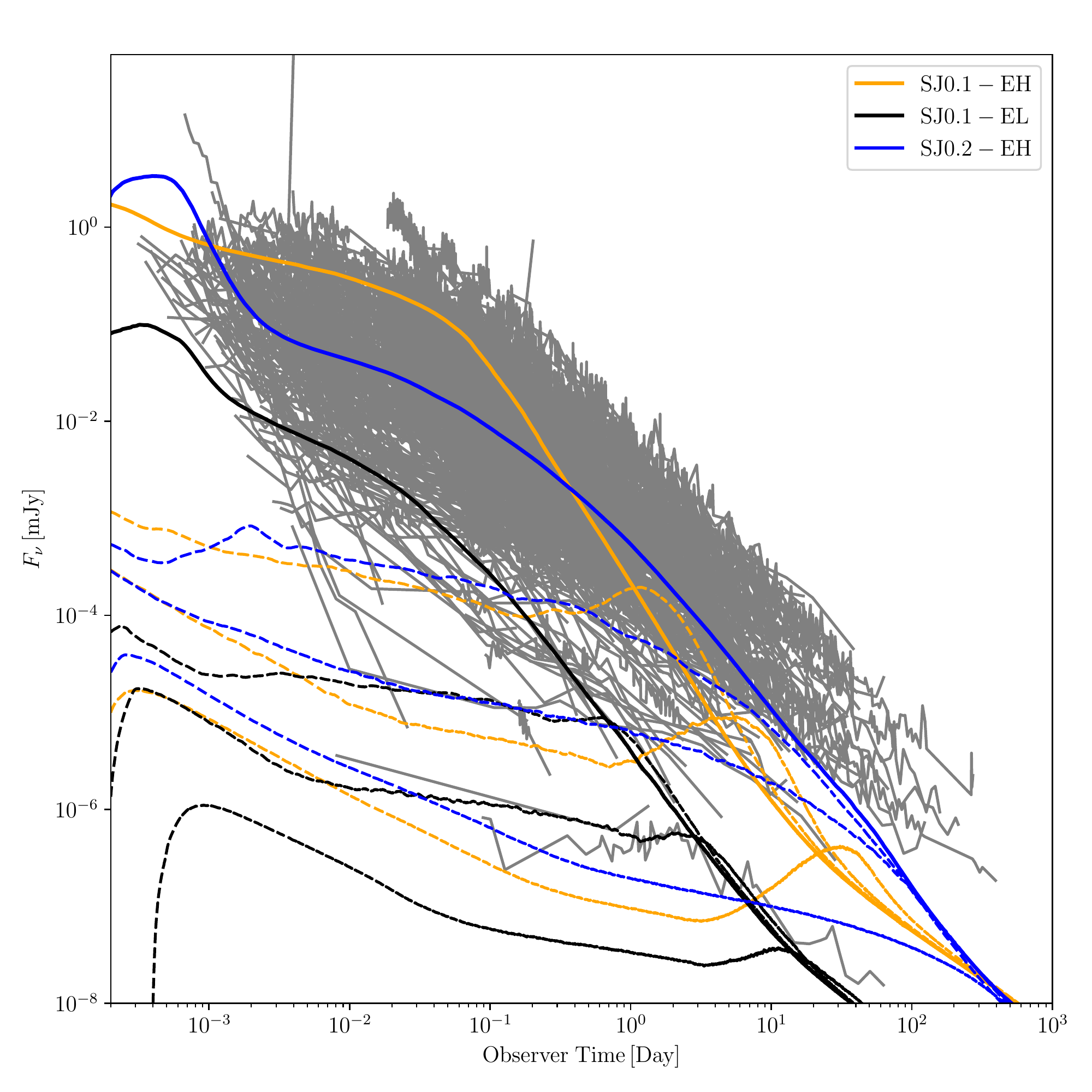}
  }
  \caption{
    The on- and off-axis multi-frequency light curves from
    FTD jet simulations. Figures \ref{fig:LC_SJ0.1-EH} - \ref{fig:LC_SJ0.2-EH} represents results from SJ0.1-EH/SJ0.1-EL/SJ0.2-EH simulation, respectively. The light curves, corresponding to 
    frequency 
    $10^{15}$ Hz (purple), and $10^{17}$ Hz (black), are included in each panel from top to
    bottom. The solid lines represent
    light curves contributed by the whole optically thin fluid elements in
    the whole simulation domain. The dashed
    lines represent flux contributed by ISM component. The dotted lines represent flux contributed by stellar component. The red dots represent R-band photometry from SN1998bw \citep{1998Natur.395..670G,2017ApJ...835...64G}, which has been rescaled to redshift 1.  Light curves from different observing angles are plotted separately. 
    Figure \ref{fig:LC_grb_bg} displays all of the X-ray ($10^{17}$ Hz) light curves presented in Figures \ref{fig:LC_SJ0.1-EH} - \ref{fig:LC_SJ0.2-EH}. The solid lines represent on-axis light curves. The dashed lines are for off-axis light curves.  GRB X-ray afterglows (0.3 keV-10 keV) detected by Swift from 2005 to 2018 Dec \citep{2007A&A...469..379E,2009MNRAS.397.1177E} are plotted as gray background. We only include the ones with confirmed redshift and rescale them to redshift 1.}
  \label{fig:star_passive}
\end{figure*}

The FTD jet simulations reveal that jets emerging from the stellar progenitor have characteristic structure that differs from top-hat jet models.
We expect the light curve from realistic jets will differ from top-hat jet models as well. In Figure \ref{fig:star_passive}, we plot multi-frequency on- and off-axis synchrotron light curves calculated from the optically thin regions of the simulation domain. The micro-physical radiation parameters are listed in Table
\ref{tab:sync}. In this study, we do not consider synchrotron self-absorption and focus our attention on optical and X-ray emission. We present light curves that cover a wide range of time from the order of seconds to the order of years. For comparison, we also include the scaled R-band light curve of supernova SN1998bw \citep{1998Natur.395..670G,2017ApJ...835...64G}. For a typical GRB-SN, there are two major components (1) the afterglow (AG), which is associated with the GRB event, (2) the supernova (SN). Clear SN bumps are observed for many GRB-SN events (for reviews, see e.g. \citealt{2006ARA&A..44..507W,2011AN....332..434M,2012grb..book..169H,2017AdAst2017E...5C}). 

\subsection{Implications for on-axis prompt emission}
\begin{figure}[!ht]
  \centering
  \includegraphics[width=1.0\columnwidth]{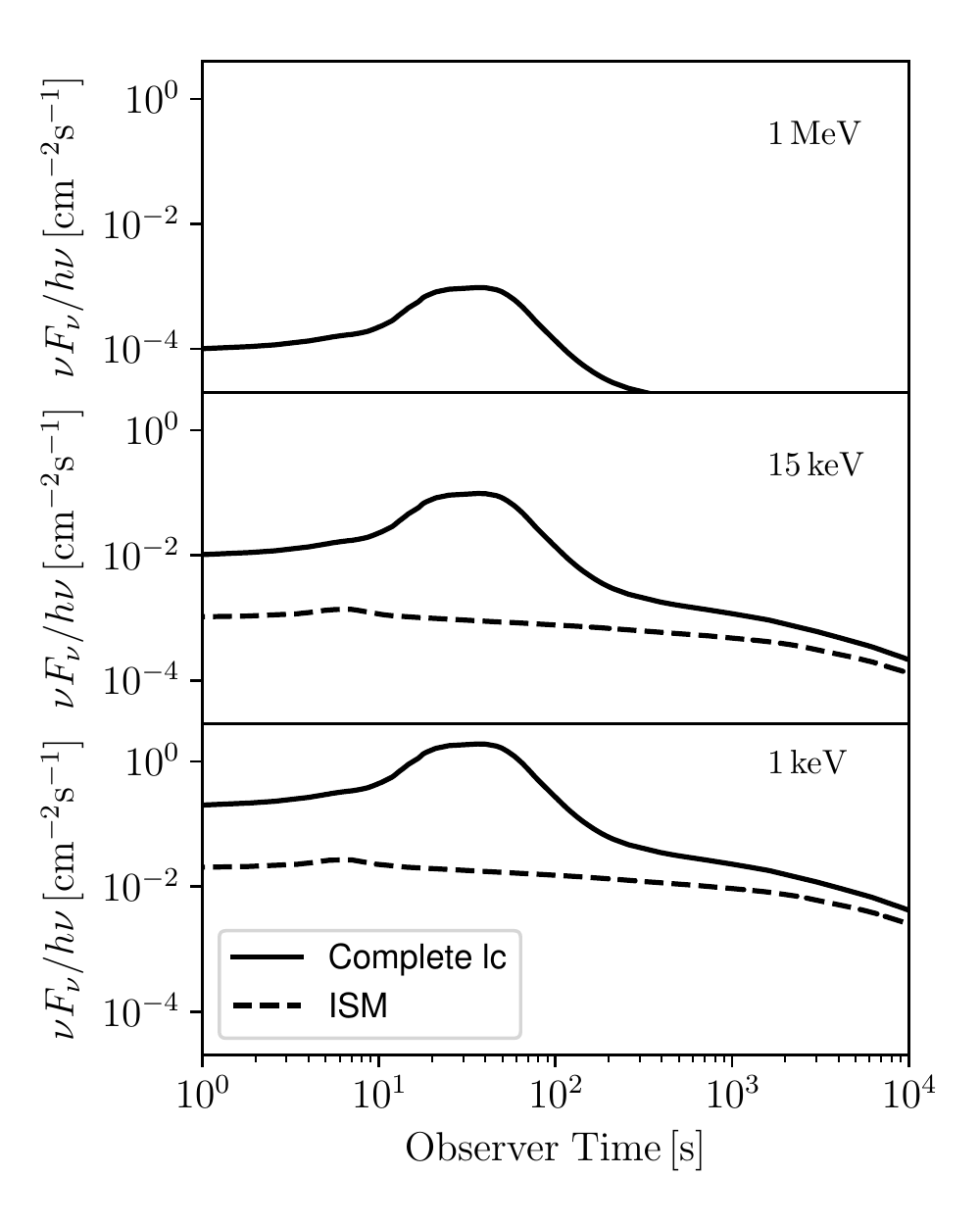}
  \caption{The initial on-axis high-frequency light curves calculated from SJ0.2-EH simulation. From top to bottom, the solid line represents the complete light curve at frequency 1MeV, 15keV, and 1keV, respectively. The emission from ISM component is shown in dashed line.}\label{fig:prompt}
\end{figure}

\begin{figure*}[t!]
  \centering
  \subfloat[\label{fig:j0.1_EH_angle}]{
    \includegraphics[clip,width=1.0\columnwidth]
    {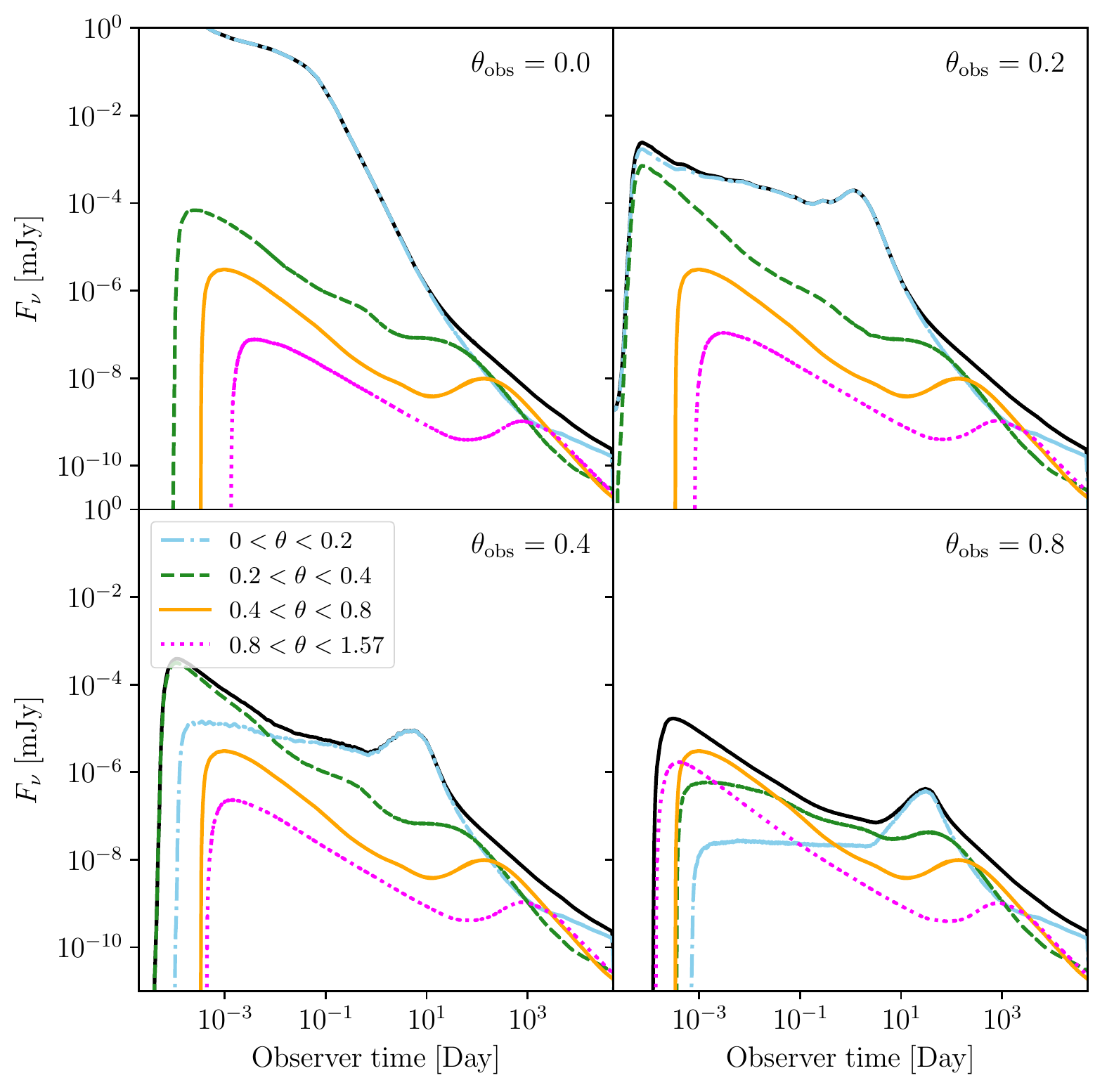} }
  \subfloat[\label{fig:j0.1_EL_angle}]{
    \includegraphics[clip,width=1.0\columnwidth]
                    {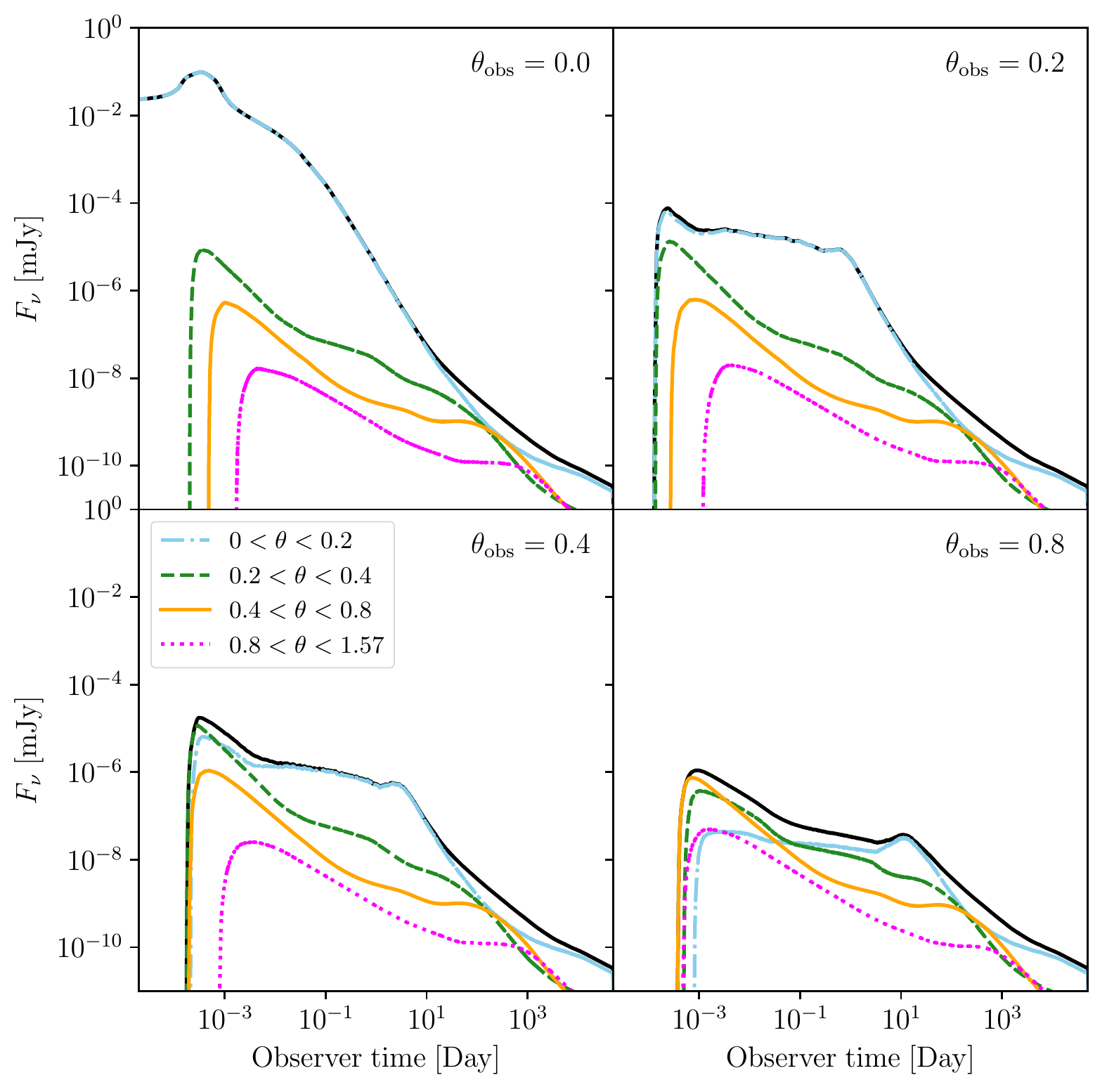}
                    }
  \caption{Angular dependent flux contribution decomposition for the on- and off-axis X-ray ($10^{17}\,\rm{Hz} $) light curves from FTD jet simulations. Figure
    \ref{fig:j0.1_EH_angle} and
    \ref{fig:j0.1_EL_angle} represent results from SJ0.1-EH simulation and SJ0.1-EL simulation, respectively. In each plot, the black solid lines display the total flux emitted by optically thin fluid elements. The emission from different angular regions in the domain is shown in different colors. The dotted-dash-blue/dashed-green/solid-orange/dotted-magenta lines show the flux contributed by fluid elements within a domain lateral angle extending from 0.0/0.2/0.4/0.8 to 0.2/0.4/0.8/1.57 [rad], respectively.}
    \label{fig:4panel_angle}
\end{figure*}

As seen in Figure \ref{fig:star_passive}, the on-axis light curves start with an early pulse followed by three major segments: an early time decay, a shallow decay, and a late steeper decay. These light curve components share similarities with canonical X-ray afterglows observed by the Swift X-ray Telescope (XRT) \citep{2009ARA&A..47..567G,2015PhR...561....1K}. The time scale of the early pulse is tens of seconds (see Figure \ref{fig:prompt} for an example), and falls within the duration of observed LGRB prompt emission. The light curve decomposition shows that the on-axis multi-frequency light curves from shock-heated ISM (dashed lines) are flat for the entire duration of prompt and shallow decay phases. We thus find the pulse mainly originates from shock-heated stellar material. Long and temporally smooth GRBs with typical variability timescales larger than a few seconds are observed and sometimes considered to arise from an external shock \citep{2016ApJ...822...63B,2018ApJ...859..163H}.

On-axis light curves from FTD simulations provides hints that, while an unified external shock model can explain both prompt and afterglow emissions, the prompt $\gamma$-ray emission of observed single pulsed GRBs may still come from shock-heated stellar materials instead of freshly shock-heated ISM materials (e.g. \citealt{2018arXiv181006965B}). Noted that, GRB prompt emission involve more complicated physics (photospheric emission, magnetic reconnection for example). Here we simply give the implication based on the analysis of hydrodynamic results.

\subsection{Implications for off-axis afterglow radiation}

The off-axis light curves (seen in Figure \ref{fig:star_passive}) exhibit clear temporal breaks as well. The break time
depends on the viewing angle. \citealt{2015ApJS..219....9W} fit the broken power-law (BPL) model to optical and X-ray light curves of 85 GRBs, and find that the break times range from a few $10^2\,\rm{s}$ to $10^3\,\rm{day}$ after
the GRB prompt emission. The break times seen in on-axis light curves in Figure \ref{fig:star_passive} fall in this range.
For off-axis light curves, an achromatic re-brightening component may appear around the break time and is then followed by a steeper decay. This is what happens for off-axis light curves from narrow jet simulations SJ0.1-EH and SJ0.1-EL, but not from the wider jet simulation SJ0.2-EH. Re-brightening features occur in observations of
long GRBs (e.g., GRB070311 \citealt{2007A&A...474..793G}, GRB081028 \citealt{2010MNRAS.402...46M}, GRB100814A \citealt{2015MNRAS.449.1024D}, GRB120326A \citealt{2014A&A...572A..55M,2014ApJ...785..113H} ), short GRBs (e.g., GRB050724 \citealt{2006A&A...454..113C},
GRB080503 \citealt{2015ApJ...807..163G}), and X-ray flashes (e.g., XRF030723 \citealt{2004ApJ...605..300H}). Various mechanisms have been proposed to explain these rebrightenings. Here we list three: the density jump model \citep{2002A&A...396L...5L,2003ApJ...591L..21D,2005NewA...10..535T,2014ApJ...789...39U,2014ApJ...792...31G}, the refreshed-shock or energy injection model \citep{1998A&A...333L..87D,1998ApJ...496L...1R,2000ApJ...532..286K,2000ApJ...535L..33S,2001ApJ...552L..35Z,2002ApJ...566..712Z,2006MNRAS.366L..13G,2006ApJ...642..354Z,2011A&A...526A.121D,2012ApJ...761..147U,2015ApJ...814....1L}, and the two-component jet model \citep{2003Natur.426..154B,2004ApJ...605..300H,2006ApJ...637..873H}, Other models can be found in e.g. \citealt{2010MNRAS.402..409K}. The FTD jet simulations demonstrate that structured jets can naturally drive the re-brightening afterglow component for off-axis observers. The ratio of its temporal width to its peak time is $\Delta T/T\sim 1$. This feature is clearly different from X-ray flares which characterize short rise time $\delta T/T<< 1$ (e.g. \citealt{2005MNRAS.364L..42F,2005Sci...309.1833B,2006ApJ...642..354Z,2006ApJ...646..351L}). Before the re-brightening, the flux decays as segments of power-law. This early decaying component for off-axis light curves originates from the shocked ISM which is different from the rapidly fading ``merger flash'' -- the non-thermal cooling emission of shock-heated merger ejecta and jet engine materials propagating in a low density environment. The magnitude of the early X-ray ``merger flash'' for GRB170817A has been estimated to lie below the instrument-detection limit of Swift when its X-ray Telescope made the first observation of the merger site \citep{2018ApJ...863...58X}. The comparison of the shape of off-axis light curves presented here and those in \citealt{2018ApJ...863...58X} shows that even though structured jets are produced in both scenarios, different set-ups of ISM density and progenitor profile can drive distinct off-axis light curves. In the work of \citealt{2018ApJ...863...58X}, the late off-axis afterglow light curves keep increasing before the external shock decelerates. In this new piece of work, we adopts a dense wind ISM profile. The off-axis external shock is decelerating after breaking out of the photosphere (revealed in the decreased value of Lorentz factor in Figures \ref{fig:dynamics_B} and \ref{fig:structure}. The off-axis light curves in this case decay with time in the beginning. The late afterglow emission (including the re-brightening components) may be overshadowed by on-going supernova emission (see Figure \ref{fig:star_passive}).

The long time monitoring of Type Ib/c SNe do not present evidence for a steeply rising light curve \citep{2006ApJ...638..930S,2014MNRAS.440..821B,2014PASA...31...22G,2018ApJ...863...32D}. 
However, broad-lined Type Ibc supernovae (SNIbc-BL), including those lacking prompt GRB emission, may be producing off-axis afterglow components which, if disentangled from supernova emission, would indicate that they harbor off-axis GRBs (see e.g. \citealt{2019arXiv190100872M}).

\subsection{light curve decomposition}

\begin{figure*}[t!]
  \centering
  \subfloat[\label{fig:j0.1_EH_radial}]{
    \includegraphics[clip,width=1.0\columnwidth]
    {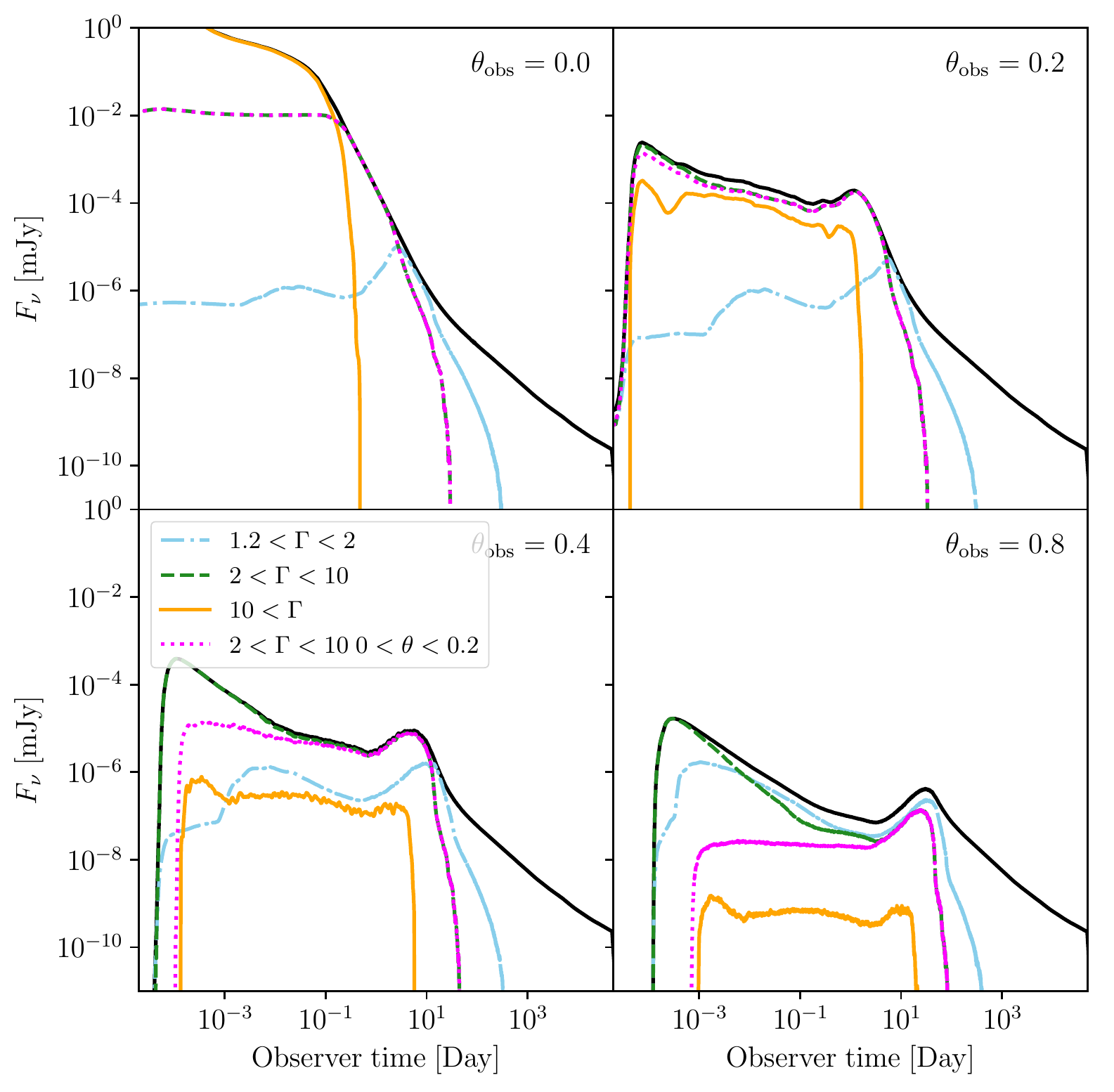} }
  \subfloat[\label{fig:j0.1_EL_radial}]{
    \includegraphics[clip,width=1.0\columnwidth]
                    {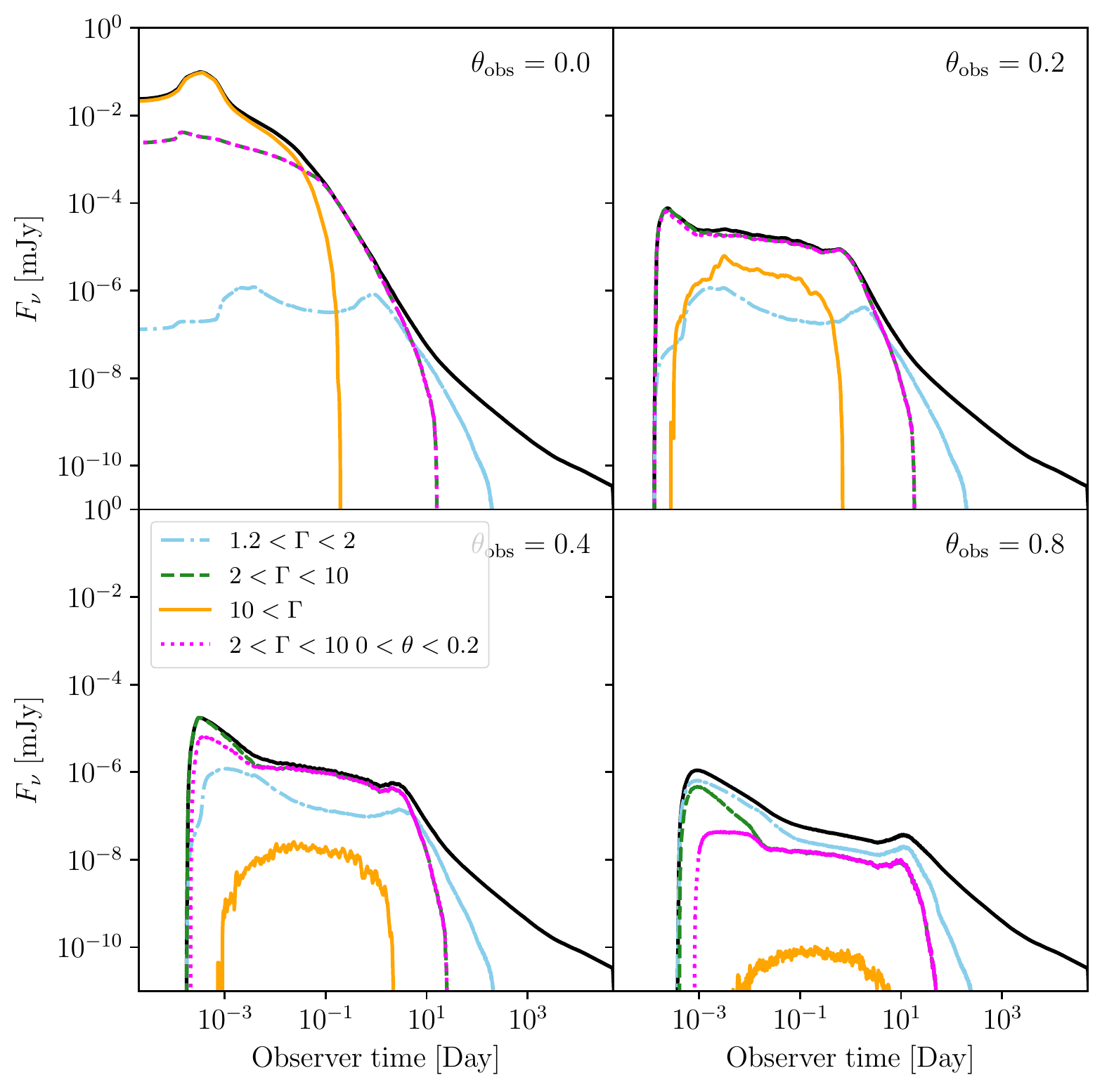}
                    }
  \caption{ Lorentz factor dependent flux contribution decomposition for the on- and off-axis X-ray ($10^{17}\,\rm{Hz} $) light curves from FTD jet simulations. Figure
    \ref{fig:j0.1_EH_radial} and
    \ref{fig:j0.1_EL_radial} represent results from SJ0.1-EH simulation and SJ0.1-EL simulation, respectively. In each plot, the black solid lines display the total flux emitted by optically thin material from the whole domain. The emission from materials with different Lorentz factor in the domain is shown in different colors. The dotted-dash-blue/dashed-green/solid-orange lines show the flux contributed by fluid elements with Lorentz factor extending from 1.2/2/10 to 2/10/maximum, respectively. The dotted magenta lines show the flux contributed by fluid elements confined within a lateral angle 0.2 and has Lorentz factor extending from 2 to 10. 
  }
    \label{fig:4panel_radial}
\end{figure*}

To better interprete the features of light curves from FTD simulations, we decompose the on- and off-axis light curves based on the lateral angle and Lorentz factor distribution of emitting materials in the shell (shown in Figure \ref{fig:4panel_angle} and Figure \ref{fig:4panel_radial}). For on-axis light curves, the materials confined within a lateral angle 0.2 determines the light curve shape for the first $\sim 10^1\,\rm{days}$, covering the prompt to early normal decay phases. The flattening part of late normal decay ($T>10\,\rm{days}$) mainly comes from high latitude emission ($\theta_{\rm{obs}}>0.2$) (see Figure \ref{fig:4panel_angle}). The first $\sim 0.1\,\rm{day}$ light curve, especially the early pulse, are emitted by materials with high Lorentz factor ($\Gamma > 10$), while the flattening part of late normal decay comes from sub-relativistic materials with $\Gamma <1.2$ (see Figure \ref{fig:4panel_radial}). For off-axis light curves, the early decay part originates from the angular region that is close to the observers' light of sight. For example, at the off-axis viewing angle $\theta_{\rm{obs}}=0.4$, the materials within the angular region $0.2<\theta<0.4$ drives the early decay. As time goes on, the emission from the central region $0.0<\theta<0.2$ gradually becomes important. It flattens the light curve and eventually drives the re-brightening components (see off-axis light curves from SJ0.1-EH and SJ0.1-EL simulations). After re-brightening, the off-axis light curves go through a steeper decay, followed by a later flattening part. The sub-relativistic materials with $\Gamma<1.2$ again drive the late flattening part (see Figure \ref{fig:4panel_radial}). Note that, generally speaking, the material in the shell expands laterally and slows down during its propagation. It is almost certain that the intially ultra-relativistic materials ($\Gamma>10$) within the central region will first contribute to the early on-axis pulsed emission and then contribute to the off-axis rebrightening components as it slows down to intermediate Lorentz factor region $2<\Gamma<10$ (see solid-orange and dotted-magenta lines in Figure \ref{fig:4panel_radial}).


\subsection{Implications for orphan afterglows}
\begin{figure}[!ht]
  \centering
  \subfloat[\label{fig:SJet_dynamics_energy}]{
    \includegraphics[clip,width=1\columnwidth]
                    {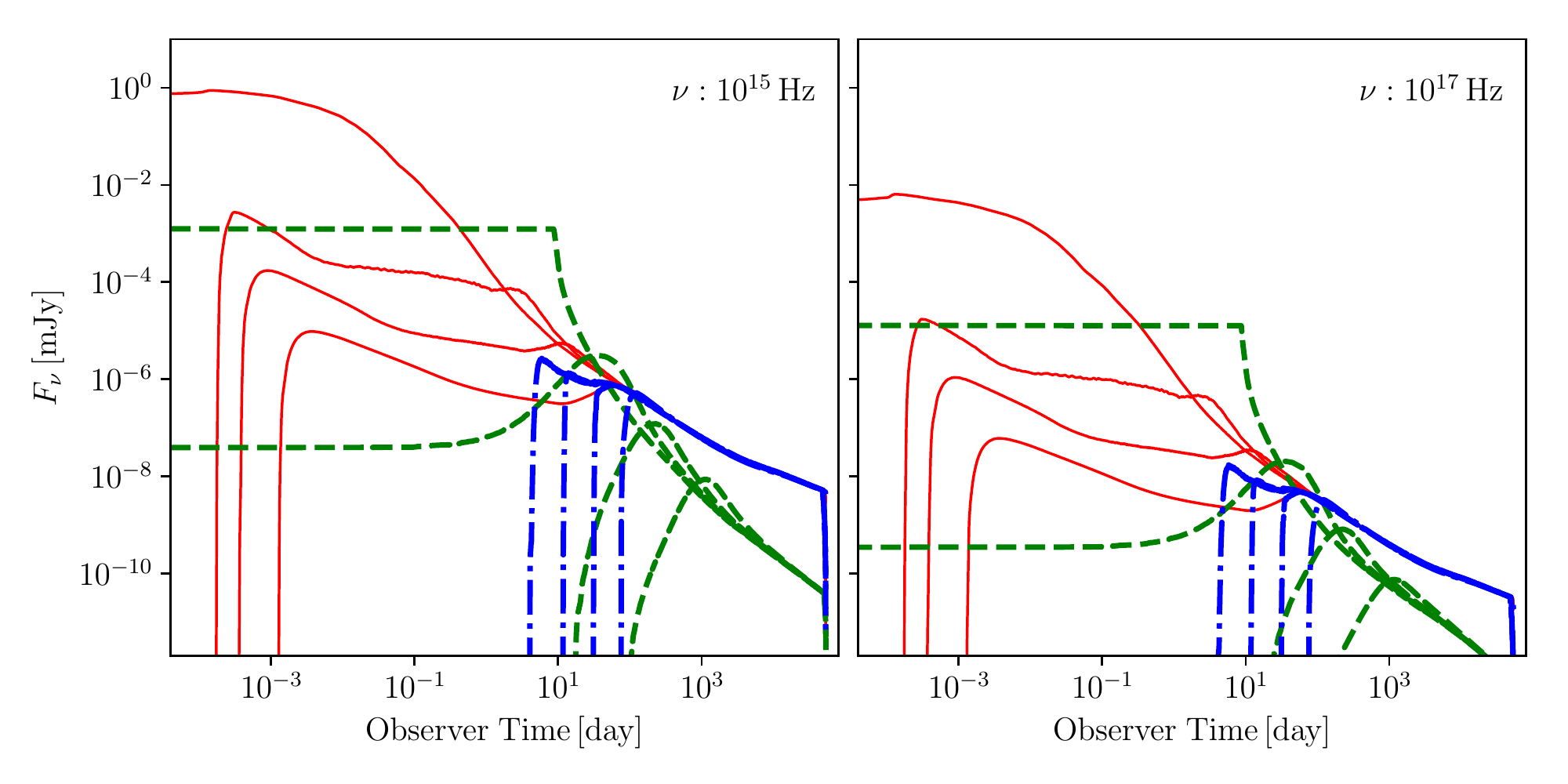}}
  \\[-0.5ex]
  \subfloat[\label{fig:SJet_dynamics_passive}]{
    \includegraphics[clip,width=1\columnwidth]
                    {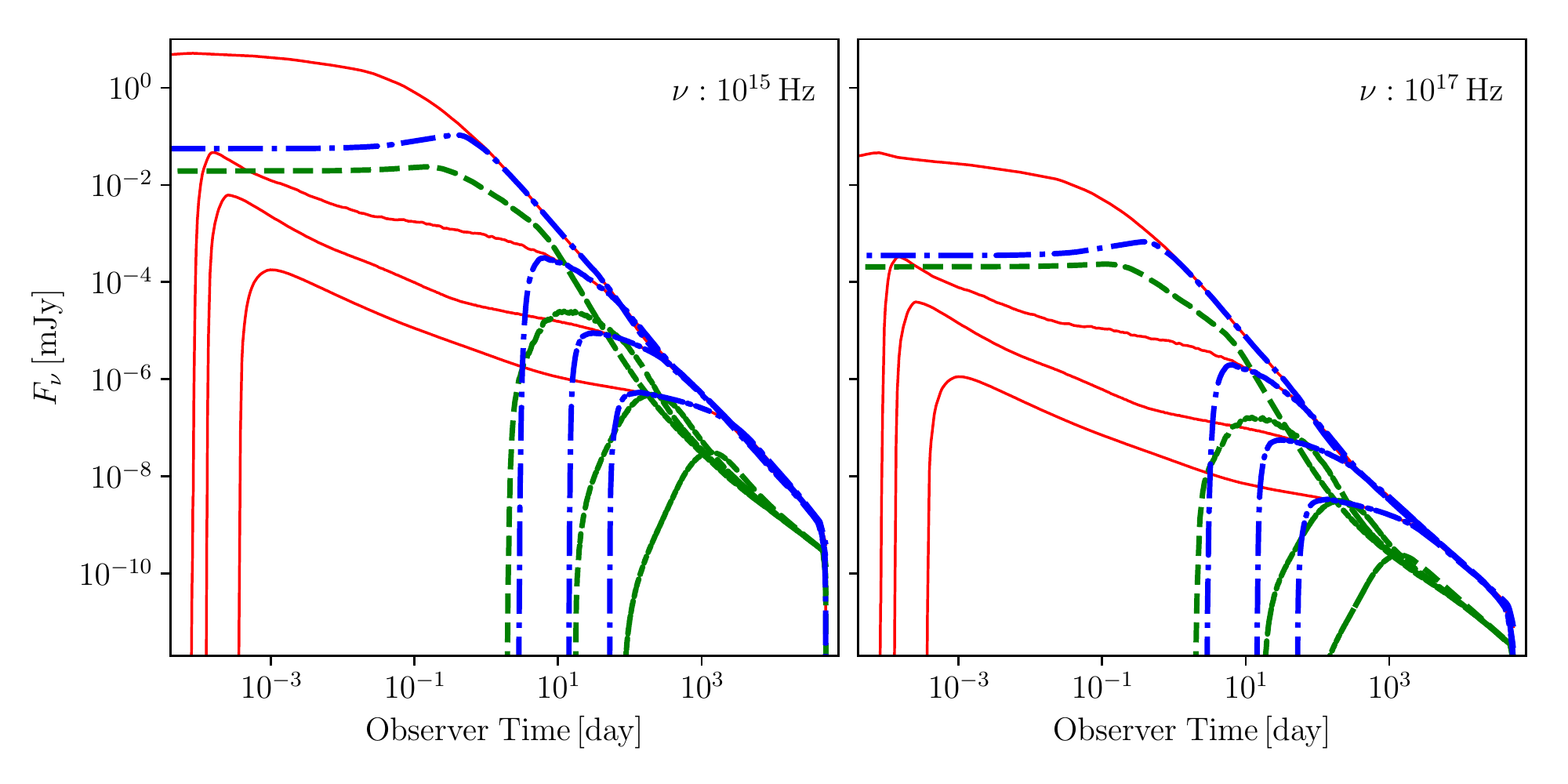}}
  \caption{ The on- and off-axis light curve comparison between FTD jet simulations and top-hat BM simulations.
    The top panel shows the comparison between the SJ0.1-EL and BM-J0.1-G100 simulations.
    The bottom  panel shows the comparison between the SJ0.2-EH and BM-J0.2-G100 simulations. Red solid
    lines represent the complete light curves calculated from FTD simulation data. Blue dotted-dashed lines represent the light curves emitted during lab time period  $t>4.37\times 10^6\,\rm{s}$. Note that we only include emission from the ISM materials here.  Green dashed lines
    represent light curves from BM simulations. The initial time for both BM 
    simulations is $t=4.37\times 10^6\,\rm{s}$. Light curves from different viewing angle
    $\theta=0.0,0.4,0.8,1.57$ are plotted from top to bottom in each panel. Different panel
    represents light curves at a different frequency: $10^{15}\ \rm{Hz}$ (on the left), $10^{17}\ \rm{Hz}$ (on the 
    right). When we only consider emission starting from
    $t=4.37\times 10^6\,\rm{s}$, the light curves from FTD simulations share common feature with BM simulations: the off-axis light curves
  exhibit a late rise-up feature. However, if we include the contribution
  from the time period $t<4.37\times 10^6\,\rm{s}$. The off-axis light curves from FTD simulations rise up very early. } \label{fig:complete_compare}
\end{figure}

The off-axis light curves in Figure \ref{fig:star_passive} rise very early on. This differs significantly from the prediction of late rise-ups in top-hat BM jet models. In Figure \ref{fig:complete_compare}, we compare the on- and off-axis light curves among three scenarios for two cases: (1) light curves (red solid line) calculated from FTD simulations -- case 1: SJ0.1-EL (on the top) and case 2: SJ0.2-EH (on the bottom). (2) light curves calculated from part of the FTD simulations covering lab frame time period $t>4.37\times 10^6\,\rm{s}$ (blue
dotted-dashed line). (3) light curves calculated from
top-hat BM simulations (green dashed line) with an initial Lorentz factor of 100, jet half opening angle 0.1 (on the top; BM-J0.1-G100 model) and 0.2 (on the bottom; BM-J0.2-G100 model). All of the light curves are calculated with the same radiation microphysical parameters as listed in Table \ref{tab:sync}. Scenario (2) and (3) use the same initial lab frame time $t=4.37 \times 10^6\,\rm{s}$ to start the calculation of synchrotron emission. The off-axis light curves from these two scenarios display qualitatively similar pattern. They all rise on a time scale of days. The rise time depends on the viewing angle. More profound difference occurs between Scenario (1) and Scenario (3). First of all, we see that off-axis light curves from Scenario (1) rise up almost instantaneously -- on a timescale of seconds up to a few minutes. There are two major reasons for this early emission feature. First of all, the afterglow emission from FTD simulations begins at a time much earlier than typical BM time scales. In FTD jet simulations, the shock front begins to surpass the photo-sphere at the radius  $r_{\rm{ph}}\sim  10^{13}\,\rm{cm}$, and starts to emit observable synchrotron radiation. It's much smaller than the initial position of typical top-hat BM models $\sim 10^{17}\rm{cm}$. 
Second, the emerged jet is structured and has a mildly relativistic
sheath extending to large lateral angle. The emission driven by the relativistic sheath is not accounted for in the top-hat jet models. Based on the above analysis, we point out that the afterglow emission calculated from top-hat BM jet models has missing  components in time and space. 

Here we revisit the concept of ``orphan afterglow'':
the observation of a late rising afterglow light curve without the accompany of prompt $\gamma$-ray burst for off-axis observers. Specific surveys have been designed and performed to search for OAs in X-ray (e.g. \citealt{1999ApJ...510..710G,2000A&A...353..998G}), Optical (e.g. \citealt{2006A&A...449...79R,2007A&A...464L..29M}) and in the radio band (e.g. \citealt{2002ApJ...576..923L,2005ApJ...631.1032R,2006ApJ...639..331G,2011MNRAS.412..634B,2011MNRAS.415....2B,2012ApJ...747...70F}). No OAs have been conclusively detected so far \citep{2014PASA...31...22G,2015A&A...578A..71G}. There are, however, OA candidates that attract attention (e.g. \citealt{2018ApJ...866L..22L}).

The off-axis afterglow light curves presented in this work advocate new templates for use in the search for off-axis afterglows or OAs. The overall shape of the off-axis light curves from FTD simulations are joint results of relativistic beaming and hydrodynamical effects.

\section{Conclusions}\label{sec:discussion}

We have presented full-time-domain (FTD) simulations of relativistic jets launched from a progenitor star using the moving-mesh hydrodynamics code \texttt{JET} \citep{2013ApJ...775...87D}. We have analyzed the angular structure of the jets at a series of fiducial times after the jet has emerged from the stellar surface and entered the afterglow stage. We find that the angular structure fits well with the universal structured jet model with angular slopes ($\alpha \equiv 6-9$), steeper than typically considered.
We calculate synchrotron light curves from the full-time-domain simulations which include early emission phase of structured jets after breakout from the photosphere. For on-axis observers we find emission components from shock-heated stellar material which may be related to single-pulsed GRB emission. We also find that the shape of calculated on-axis light curves is similar to the observed pattern of GRB afterglow light curves, featuring a steep decay, followed by a shallow phase then a second decay.

For off-axis observers, we find that the light curves rise earlier than previously expected, even for observers at large viewing angles.  This early rising is different from the late rising predictions derived from top-hat Blandford-McKee jet models, and is consistent with the fact that no ``orphan afterglow'' based on those models has been conclusively detected so far. Improved afterglow templates based on full-time-domain simulations may thus be helpful for orphan or off-axis afterglow searches. Such off-axis light curve  templates, generally speaking, feature a short-period initial decay which originates from the part of the shell that moves toward the observer, followed by a period of flattening from the decelerating upper shell (sometimes with re-brightening components) and a late steeper decay.

We have found that off-axis light curves sometimes display late rebrightenings due to the relativistic core of the jet decelerating and emitting into off-axis directions. These late emission components may be observable but can be mixed with, or hidden by, supernova emission. Broad-lined Type Ibc supernovae (Sne Ic-bl), including those lacking prompt GRB emission, may be producing these off-axis afterglow components. if they can be disentangled from supernova emission, it would indicate that these Sne Ic-bl supernova harbor off-axis GRBs. Future studies, combining the computation of viewing angle dependant Sne Ic-bl emission \citep{2018ApJ...860...38B} and synchrotron afterglows, and with their comparison to observations, should help us better understand the nature of GRB-supernovae.

The results presented in this study may be helpful for sky surveys searching for off-axis and orphan afterglows. Nevertheless, our current conclusions are based on a specific stellar progenitor profile and simplified jet engine models.  One of the caveats in our study is the insufficient exploration of parameter space. To better understand the dynamics and radiation of long gamma-ray burst jets, more full-time-domain hydrodynamic simulations and radiation modeling are needed.

We appreciate helpful discussions with  Brian Metzger, Raffaella Margutti,
Maryam Modjaz, Andrei Gruzinov, Kate Alexander, Paz Beniamini, Paul Duffell, and Yiyang Wu. We thank George Wong for providing the
visualization tool tailored to visualize checkpoints produced by
JET simulations. This work made use of data supplied by the UK Swift Science Data Centre at the University of Leicester.

\software{\texttt{JET} \citep{2013ApJ...775...87D}}

\bibliography{ms.bib}


\listofchanges

\end{document}